\documentclass[twocolumn,times,tighten]{aastex631}

\usepackage{graphicx} 
\usepackage{amsmath}
\usepackage{xcolor}
\usepackage{mathtools}
\usepackage{hyperref}
\usepackage{array}
\usepackage{longtable}

\usepackage{pifont}

\usepackage{amsmath}	
\usepackage{amssymb}	
\usepackage{amsbsy}	
\usepackage{newtxtt,newtxmath}
\usepackage{catchfile}
\usepackage{enumitem}

\newcommand{\msol}{{\rm\,M$_\odot$}}

\newcommand{\oii}{[O{\sc ii}]}
\newcommand{\oiii}{[O{\sc iii}]}
\newcommand{\lya}{\rm{Ly}$\alpha$}
\newcommand{\halpha}{\rm{H}$\alpha$}
\newcommand{\hbeta}{\rm{H}$\beta$}
\newcommand{\qf}{$Q_f$}
\newcommand{\cigale}{\texttt{CIGALE}}
\newcommand{\lephare}{\texttt{LePHARE}}

\newcommand{\uniqueSources}{266,284} 
\newcommand{\uniqueReliableSources}{184,633} 
\newcommand{\uniqueBLSources}{4,338} 

\newcommand{\allSources}{487,666} 
\newcommand{\allreliableSources}{288,044} 
\newcommand{\allreliableBLSources}{10,505} 

\newcommand{\nprograms}{$138$} 
\newcommand{\nprogramsOPT}{$71$} 
\newcommand{\nprogramsNIR}{$64$} 
\newcommand{\nprogramsMIR}{$1$} 
\newcommand{\nprogramsRADIO}{$20$} 
\newcommand{\nprogramsGROUND}{$127$} 
\newcommand{\nprogramsSPACE}{$11$} 

\newcommand{\NGroups}{124,211} 
\newcommand{\AvgGroupSize}{3} 
\newcommand{\MaxGroupSize}{25} 

\usepackage{lipsum}
\begin{document}
\title{COSMOS Spectroscopic Redshift Compilation (First Data Release):\\ 488k Redshifts Encompassing Two Decades of Spectroscopy}

\received{}
\revised{}
\accepted{}

\submitjournal{ApJ}

\shorttitle{COSMOS Spec-$z$ Compilation}
\shortauthors{Khostovan et al.}

\correspondingauthor{Ali Ahmad Khostovan}
\email{akhostov@gmail.com}

\author[0000-0002-0101-336X]{Ali Ahmad Khostovan}
\affiliation{Department of Physics and Astronomy, University of Kentucky, 505 Rose Street, Lexington, KY 40506, USA}
\affiliation{Laboratory for Multiwavelength Astrophysics, School of Physics and Astronomy, Rochester Institute of Technology, 84 Lomb Memorial Drive, Rochester, NY 14623, USA}

\author[0000-0001-9187-3605]{Jeyhan S. Kartaltepe}
\affiliation{Laboratory for Multiwavelength Astrophysics, School of Physics and Astronomy, Rochester Institute of Technology, 84 Lomb Memorial Drive, Rochester, NY 14623, USA}

\author[0000-0001-7116-9303]{Mara Salvato}
\affiliation{Max-Planck-Institut f\"ur extraterrestrische Physik, Giessenbachstrasse 1, 85748 Garching, Germany}

\author[0000-0002-7303-4397]{Olivier Ilbert}
\affiliation{Aix Marseille Univ, CNRS, CNES, LAM, Marseille, France}

\author[0000-0002-0930-6466]{Caitlin M. Casey}
\affiliation{Department of Physics, University of California, Santa Barbara, Santa Barbara, CA 93109, USA}
\affiliation{Department of Astronomy, The University of Texas at Austin, 2515 Speedway Blvd Stop C1400, Austin, TX 78712, USA}
\affiliation{Cosmic Dawn Center (DAWN), Denmark}

\author[0000-0002-4205-9567]{Hiddo Algera}
\affiliation{Institute of Astronomy and Astrophysics, Academia Sinica, 11F of Astronomy-Mathematics Building, No.1, Sec. 4, Roosevelt Rd, Taipei 106216, Taiwan, R.O.C.}

\author[0000-0002-0243-6575]{Jacqueline Antwi-Danso}
\affiliation{David A. Dunlap Department of Astronomy and Astrophysics, University of Toronto, 50 St. George Street, Toronto, Ontario, M5S}

\author[0000-0003-4569-2285]{Andrew Battisti}
\affiliation{International Centre for Radio Astronomy Research, University of Western Australia, 35 Stirling Highway, Crawley 26WA 6009, Australia}
\affiliation{ARC Centre of Excellence for All Sky Astrophysics in 3 Dimensions (ASTRO 3D), Australia}
\affiliation{Research School of Astronomy and Astrophysics, Australian National University, Cotter Road, Weston Creek, ACT 2611, Australia}

\author[0000-0002-0245-6365]{Malte Brinch}
\affiliation{Instituto de F\'isica y Astronom\'ia, Universidad de Valpara\'iso, Avda. Gran Bretana~ 1111, Valpara\'iso, Chile}

\author[0000-0002-5059-6848]{Marcella Brusa}
\affiliation{Dipartimento di Fisica e Astronomia ‘Augusto Righi’, Alma Mater Studiorum Università di Bologna, via Gobetti 93/2,
40129 Bologna, Italy}
\affiliation{INAF-Osservatorio di Astrofisica e Scienza dello Spazio di Bologna, via Gobetti 93/3, 40129 Bologna, Italy}

\author[0000-0003-2536-1614]{Antonello Calabr\`o}
\affiliation{INAF—Osservatorio Astronomico di Roma, via di Frascati 33, 00078 Monte Porzio Catone, Italy}

\author[0000-0003-3578-6843]{Peter L.~Capak}
\affiliation{Cosmic Dawn Center (DAWN), Denmark}
\affiliation{Niels Bohr Institute, University of Copenhagen, Jagtvej 128, DK-2200, Copenhagen, Denmark}

\author[0000-0003-3691-937X]{Nima Chartab}
\affiliation{Caltech/IPAC, MS 314-6, 1200 E. California Blvd. Pasadena, CA 91125, USA}

\author[0000-0003-3881-1397]{Olivia R. Cooper}\altaffiliation{NSF Graduate Research Fellow}
\affiliation{Department of Astronomy, The University of Texas at Austin, 2515 Speedway Blvd Stop C1400, Austin, TX 78712, USA}

\author[0000-0002-1803-794X]{Isa G. Cox}
\affiliation{Laboratory for Multiwavelength Astrophysics, School of Physics and Astronomy, Rochester Institute of Technology, 84 Lomb Memorial Drive, Rochester, NY 14623, USA}

\author[0000-0003-4919-9017]{Behnam Darvish}
\affiliation{Department of Physics and Astronomy, University of California, Riverside, 900 University Avenue, Riverside, CA 92521, USA}

\author[0000-0003-4761-2197]{Nicole E.~Drakos}
\affiliation{Department of Physics and Astronomy, University of Hawaii, Hilo, 200 W. Kawili St., Hilo, HI 96720, USA}

\author[0000-0002-9382-9832]{Andreas L. Faisst}
\affiliation{Caltech/IPAC, MS 314-6, 1200 E. California Blvd. Pasadena, CA 91125, USA}

\author{Matthew R.~George}
\affiliation{Department of Astronomy, University of California, Berkeley, CA 94720, USA}
\affiliation{Lawrence Berkeley National Laboratory, 1 Cyclotron Rd., Berkeley, CA, 94720, USA}

\author[0000-0002-0236-919X]{Ghassem Gozaliasl}
\affiliation{Department of Computer Science, Aalto University, P.O. Box 15400, FI-00076 Espoo, Finland}
\affiliation{Department of Physics, University of Helsinki, P.O. Box 64, FI-00014 Helsinki, Finland}

\author[0000-0003-0129-2079]{Santosh Harish}
\affiliation{Laboratory for Multiwavelength Astrophysics, School of Physics and Astronomy, Rochester Institute of Technology, 84 Lomb Memorial Drive, Rochester, NY 14623, USA}

\author[0000-0002-0797-0646]{G\"unther Hasinger}
\affiliation{TU Dresden, Institute of Nuclear and Particle Physics, 01062 Dresden, Germany}
\affiliation{DESY, Notkestra{\ss}e 85, 22607 Hamburg, Germany}
\affiliation{Deutsches Zentrum f\"ur Astrophysik, Postplatz 1, 02826 G\"orlitz, Germany}

\author[0009-0007-3673-4523]{Hossein Hatamnia}
\affiliation{Department of Physics and Astronomy, University of California, Riverside, 900 University Avenue, Riverside, CA 92521, USA}

\author[0000-0001-6958-0304]{Angela Iovino}
\affiliation{INAF – Osservatorio Astronomico di Brera, via Brera 28, I-20121 Milano, Italy}

\author[0000-0002-8412-7951]{Shuowen Jin}
\affiliation{Cosmic Dawn Center (DAWN), Denmark}
\affiliation{DTU Space, Technical University of Denmark, Elektrovej 327, DK2800 Kgs. Lyngby, Denmark}

\author[0000-0001-9044-1747]{Daichi Kashino}
\affiliation{National Astronomical Observatory of Japan, 2-21-1 Osawa, Mitaka, Tokyo, 181-8588, Japan}

\author[0000-0002-6610-2048]{Anton M. Koekemoer}
\affiliation{Space Telescope Science Institute, 3700 San Martin Drive, Baltimore, MD 21218, USA}

\author[0000-0002-0322-6131]{Ronaldo Laishram}
\affiliation{National Astronomical Observatory of Japan, 2-21-1 Osawa, Mitaka, Tokyo, 181-8588, Japan}

\author[0000-0001-9299-5719]{Khee-Gan Lee}
\affiliation{Kavli IPMU (WPI), UTIAS, The University of Tokyo, Kashiwa, Chiba 277-8583, Japan}
\affiliation{Center for Data-Driven Discovery, Kavli IPMU (WPI), UTIAS, The University of Tokyo, Kashiwa, Chiba 277-8583, Japan}

\author[0000-0002-3535-4066]{Jitrapon Lertprasertpong}
\affiliation{Laboratory for Multiwavelength Astrophysics, School of Physics and Astronomy, Rochester Institute of Technology, 84 Lomb Memorial Drive, Rochester, NY 14623, USA}

\author[0000-0002-6423-3597]{Simon J.~Lilly}
\affiliation{Department of Physics, ETH Z\"urich, Wolfgang-Pauli-Strasse 27, Z\"urich, 8093, Switzerland}

\author[0000-0001-9773-7479]{Daizhong Liu}
\affiliation{Purple Mountain Observatory, Chinese Academy of Sciences, 10 Yuanhua Road, Nanjing 210023, China}

\author[0000-0001-5382-6138]{Daniel C.~Masters}
\affiliation{Caltech/IPAC, MS 314-6, 1200 E. California Blvd. Pasadena, CA 91125, USA}

\author[0000-0001-5846-4404]{Bahram Mobasher}
\affiliation{Department of Physics and Astronomy, University of California, Riverside, 900 University Avenue, Riverside, CA 92521, USA}

\author[0000-0002-7402-5441]{Tohru Nagao}
\affiliation{Research Center for Space and Cosmic Evolution, Ehime University, Bunkyo-cho 2-5, Matsuyama, Ehime 790-8577, Japan}

\author[0000-0003-3228-7264]{Masato Onodera}
\affiliation{Department of Astronomical Science, The Graduate University for Advanced Studies, SOKENDAI, 2-21-1 Osawa, Mitaka, Tokyo 181-8588, Japan}
\affiliation{Subaru Telescope, National Astronomical Observatory of Japan, 650 North Aohoku Place, Hilo, HI, 96720, USA}

\author[0000-0003-0939-9671]{Yingjie Peng}
\affiliation{Department of Astronomy, School of Physics, Peking University, 5 Yiheyuan Road, Beijing 100871, China}
\affiliation{Kavli Institute for Astronomy and Astrophysics, Peking University, Beijing 100871, China}

\author[0000-0002-1233-9998]{David B.~Sanders}
\affiliation{Institute for Astronomy, University of Hawaii, 2680 Woodlawn Drive, Honolulu, HI 96822, USA}

\author[0000-0003-4792-9119]{Ryan L.~Sanders}
\affiliation{Department of Physics and Astronomy, University of Kentucky, 505 Rose Street, Lexington, KY 40506, USA}

\author[0000-0002-0364-1159]{Zahra Sattari}
\affiliation{Caltech/IPAC, MS 314-6, 1200 E. California Blvd. Pasadena, CA 91125, USA}

\author[0000-0002-0438-3323]{Nick Scoville}
\affiliation{Astronomy Department, California Institute of Technology, 1200 E. California Boulevard, Pasadena, CA 91125, USA}

\author[0000-0001-7811-9042]{Ekta A.~Shah}
\affiliation{Department of Physics and Astronomy, University of California, Davis, One Shields Avenue, Davis, CA 95616, USA}
\affiliation{Laboratory for Multiwavelength Astrophysics, School of Physics and Astronomy, Rochester Institute of Technology, 84 Lomb Memorial Drive, Rochester, NY 14623, USA}

\author[0000-0002-0000-6977]{John~D. Silverman}
\affiliation{Kavli IPMU (WPI), UTIAS, The University of Tokyo, Kashiwa, Chiba 277-8583, Japan}
\affiliation{Center for Data-Driven Discovery, Kavli IPMU (WPI), UTIAS, The University of Tokyo, Kashiwa, Chiba 277-8583, Japan}
\affiliation{Department of Astronomy, School of Science, The University of Tokyo, 7-3-1 Hongo, Bunkyo, Tokyo 113-0033, Japan}
\affiliation{Center for Astrophysical Sciences, Department of Physics \& Astronomy, Johns Hopkins University, Baltimore, MD 21218, USA}

\author[0000-0001-7266-930X]{Nao Suzuki}
\affiliation{Lawrence Berkeley National Laboratory, 1 Cyclotron Rd., Berkeley, CA, 94720, USA}
\affiliation{Department of Physics, Florida State University, 77 Chieftan Way, Tallahassee, FL 32306, USA}
\affiliation{Laboratoire de Physique Nucleaire et de Hautes-Energies, 4 Place Jussieu, 75005 Paris, France}

\author[0000-0003-0749-4667]{Sina Taamoli}
\affiliation{Department of Physics and Astronomy, University of California, Riverside, 900 University Avenue, Riverside, CA 92521, USA}

\author[0000-0002-5011-5178]{Masayuki Tanaka}
\affiliation{National Astronomical Observatory of Japan, 2-21-1 Osawa, Mitaka, Tokyo, 181-8588, Japan}
\affiliation{Department of Astronomical Science, The Graduate University for Advanced Studies, SOKENDAI, 2-21-1 Osawa, Mitaka, Tokyo, 181-8588, Japan}

\author{Lidia A.~M.~Tasca}
\affiliation{Aix Marseille Univ, CNRS, CNES, LAM, Marseille, France}

\author[0000-0003-3631-7176]{Sune Toft}
\affiliation{Cosmic Dawn Center (DAWN), Denmark}
\affiliation{Niels Bohr Institute, University of Copenhagen, Jagtvej 128, DK-2200, Copenhagen, Denmark}

\author[0009-0005-3133-1157]{Greta Toni}
\affiliation{Dipartimento di Fisica e Astronomia ‘Augusto Righi’, Alma Mater Studiorum Università di Bologna, via Gobetti 93/2,
40129 Bologna, Italy}
\affiliation{INAF-Osservatorio di Astrofisica e Scienza dello Spazio di Bologna, via Gobetti 93/3, 40129 Bologna, Italy}
\affiliation{Zentrum f\"{u}r Astronomie, Universit\"{a}t Heidelberg, Philosophenweg 12, D-69120, Heidelberg, Germany}

\author[0000-0002-3683-7297]{Benny Trakhtenbrot}
\affiliation{School of Physics and Astronomy, Tel Aviv University, Tel Aviv 69978, Israel}

\author[0000-0002-1410-0470]{Jonathan R.~Trump}
\affiliation{Department of Physics, 196 Auditorium Road, Unit 3046, University of Connecticut, Storrs, CT 06269, USA}

\author[0000-0002-6748-0577]{Mattia~Vaccari}
\affiliation{Inter-University Institute for Data Intensive Astronomy, Department of Astronomy, University of Cape Town, 7701 Rondebosch, Cape Town, South Africa}
\affiliation{Inter-University Institute for Data Intensive Astronomy, Department of Physics and Astronomy, University of the Western Cape, Robert Sobukwe Road, 7535 Bellville, Cape Town, South Africa}
\affiliation{INAF – Istituto di Radioastronomia, via Gobetti 101, 40129 Bologna, Italy}

\author[0000-0001-6477-4011]{Francesco Valentino}
\affiliation{Cosmic Dawn Center (DAWN), Denmark}
\affiliation{DTU Space, Technical University of Denmark, Elektrovej 327, DK-2800 Kgs. Lyngby, Denmark}

\author[0000-0002-8163-0172]{Brittany N.~Vanderhoof}
\affiliation{Space Telescope Science Institute, 3700 San Martin Drive, Baltimore, MD 21218, USA}
\affiliation{Laboratory for Multiwavelength Astrophysics, School of Physics and Astronomy, Rochester Institute of Technology, 84 Lomb Memorial Drive, Rochester, NY 14623, USA}

\author[0000-0003-1614-196X]{John R.~Weaver}
\affiliation{Department of Astronomy, University of Massachusetts, 710 North Pleasant Street, Amherst, MA 01003-9305, USA}

\author[0000-0001-7095-7543]{Min S.~Yun}
\affiliation{Department of Astronomy, University of Massachusetts, 710 North Pleasant Street, Amherst, MA 01003-9305, USA}

\author[0000-0002-7051-1100]{Jorge A. Zavala}
\affiliation{Department of Astronomy, University of Massachusetts, 710 North Pleasant Street, Amherst, MA 01003-9305, USA}

\begin{abstract}
We present the COSMOS Spectroscopic Redshift Compilation encompassing $\sim$20 years of spectroscopic redshifts within a 10 deg$^2$ area centered on the 2 deg$^2$ COSMOS legacy field. This compilation contains \allSources~redshifts of \uniqueSources~unique objects from \nprograms~individual programs up to $z\sim8$ with median stellar mass $\sim10^{8.4}$ to $10^{10}$ \msol~(redshift dependent). Rest-frame $NUVrJ$ colors and SFR -- stellar mass correlations show the compilation primarily contains low- to intermediate-mass star-forming and massive, quiescent galaxies at $z<1.25$ and mostly low-mass bursty star-forming galaxies at $z>2$. Sources in the compilation cover a diverse range of environments, including protoclusters such as ``Hyperion''. The full compilation is 50\% spectroscopically complete by $i\sim23.4$ and $K_s\sim21.6$ mag; however, this is redshift dependent. Spatially, the compilation is $>50$\% ($>30$\%) complete within the central (outer) region limited to $i<24$ mag and $K_s<22.5$ mag, separately. We demonstrate how the compilation can be used to validate photometric redshifts and investigate calibration metrics. By training self-organizing maps on COSMOS2020/Classic and projecting the compilation onto it, we find key subpopulations currently lacking spectroscopic coverage including $z<1$ intermediate-mass quiescent and low-/intermediate-mass bursty star-forming galaxies, $z\sim2$ massive quiescent galaxies, and $z>3$ massive star-forming galaxies. This highlights how combining self-organizing maps with our compilation can provide guidance for future spectroscopic observations to get a complete spectroscopic view of galaxy populations. Lastly, the compilation will undergo periodic data releases incorporating new spectroscopic redshifts and providing a lasting legacy resource for the community.
\end{abstract}

\section{Introduction}\label{sec:intro}

In the next several years, astronomy will see a large influx of both photometric and spectroscopic data sets. Planned surveys with space-based missions (e.g., \textit{Euclid}, \textit{Roman}) and ground-based facilities (e.g., DESI, Rubin, Subaru/PFS, VLT/MOONS, SDSS-V, 4MOST) will observe hundreds of millions of galaxies and AGN within the optical to near-infrared and push the limits towards fainter and rarer galaxy populations. However, at the same time there is a plethora of past spectroscopic data sets and measurements that have been collected over the past twenty years but are scattered within many archival databases (e.g., IRSA IPAC, ESO Science Portal), catalog services (e.g., CDS/VizieR), survey-specific websites, tables in peer-reviewed publications, and simply on the hard drives of the PIs (e.g., programs that have redshift measurements but were never published). Gathering, organizing, and compiling the wealth of spectroscopic redshift measurements into one uniform data product is crucial not only in regard to data management, but also in supporting current and future science and calibration initiatives and programs. 

The Cosmic Evolution Survey (COSMOS; $\alpha = 10^\textrm{h}00^\textrm{m}28.6^\textrm{s}$, $\delta = +2^\circ 12^\textrm{m}21.0^\textrm{s}$; J2000; \citealt{Scoville2007}) began as the largest contiguous area \textit{HST} extragalactic imaging survey using ACS/F814W to observe $\sim2$\,deg$^2$ \citep{Koekemoer2007} with the main focus on observing galaxy populations up to $z \sim 6$ to study galaxy formation and evolution, mapping dark matter distributions, measuring key statistical distributions relating to stellar mass and luminosity of galaxy populations, and investigating the role of AGN, black hole growth, and large-scale structure on galaxy evolution. 

Over the past 20 years, COSMOS has been established as the premier legacy field for extragalactic astrophysics with over 40 bands of photometry covering a wide multi-wavelength space from X-ray to radio. The first published COSMOS catalog (COSMOS2009; \citealt{Capak2007,Ilbert2009,Salvato2009}) presented the photometry and derived measurements for $\sim2$ million galaxies supported with multi-wavelength imaging from programs such as Subaru/Suprime-Cam COSMOS-20 \citep{Taniguchi2007,Taniguchi2015} and \textit{Spitzer} S-COSMOS \citep{Sanders2007}. Several key imaging surveys came later including but not limited to: deep CFHT $u$ imaging (CLAUDS; \citealt{Sawicki2019}), Hyper Suprime-Cam Subaru Strategic Program (HSC-SSP; \citealt{Aihara2018}), $YJHK_s$ and NB118 narrowband coverage from UltraVISTA \citep{McCracken2012}, \textit{Spitzer}/IRAC and MIPS imaging \citep{LeFloch2009,Ashby2013,Ashby2015,Ashby2018,Steinhardt2014}, deep \textit{GALEX} imaging \citep{Zamojski2007}, X-ray imaging from \textit{XMM} \citep{Cappelluti2007, Hasinger2007,Brusa2010} and \textit{Chandra} \citep{Elvis2009,Civano2012,Civano2016,Marchesi2016} further optical and near-IR \textit{HST} imaging \citep{Nayyeri2017}, far-IR \textit{Herschel} coverage \citep{Lutz2011,Oliver2012}, and sub-millimeter and radio coverage \citep{Bertoldi2007,Schinnerer2007,Scott2008,Miettinen2017,Smolcic2017,Liu2019}. The combination of all these imaging datasets led to updated catalogs: COSMOS2015 \citep{Laigle2016} and COSMOS2020 \citep{Weaver2022} resulting in $\sim 1.2$ and $1.7$ million galaxies up to $z \sim 6$ and $10$, respectively. Most recently, COSMOS now includes \textit{JWST}/NIRCam and MIRI imaging from COSMOS-Web (PIs: J.~S.~Kartaltepe \& C.~Casey, \citealt{Casey2023}) and PRIMER (PI: J.~Dunlop) providing the needed deep near/mid-IR coverage to detect and constrain the properties of high-$z$ galaxy populations.

The wealth of multi-wavelength imaging datasets available in COSMOS has also enabled a multitude of spectroscopic follow-up programs. These programs utilize a variety of both ground-based (e.g., Keck, Gemini, VLT, Subaru) and space-based (e.g., \textit{HST}) facilities. Some of the larger surveys among these have published catalogs and are also included in our compilation, including $z$COSMOS (\citealt{Lilly2007}; PI: S.~Lilly), MOSDEF \citep{Kriek2015}, FMOS-COSMOS \citep{Kashino2019}, C3R2 \citep{Masters2017,Masters2019}, 3D-\textit{HST} \citep{Brammer2012, Momcheva2016}, VUDS \citep{Tasca2017}, and LEGA-C \citep{vanderWel2021}. COSMOS is also the focus of several \textit{JWST} spectroscopic programs such as AURORA \citep{Shapley2025}, PASSAGE (PI: Matthew Malkan), and COSMOS-3D (PI: Koki Kakiichi).

Given the multitude and depth of both photometric and spectroscopic data available, COSMOS is a key focus of \textit{Euclid} as an Auxiliary Field \citep{Scaramella2022} and the Vera Rubin Observatory as a Deep Drilling Field \citep{Brandt2018} for calibration purposes. Therefore, having a spectroscopic compilation that organizes the above-mentioned spectroscopic programs as well as the many programs scattered within publications, catalog services, survey websites, and other sources is critical for supporting these facilities. Furthermore, such a large compilation is also necessary to support current science initiatives as well as help guide/design future science goals and objectives as new facilities start to achieve first light (e.g., Subaru/PFS, \textit{Roman}, VLT/MOONS).

In this paper, we present the first data release of the COSMOS Spectroscopic Redshift Compilation. 
We describe how the data were collected in \S\ref{sec:data} and summarize each program included in the compilation. We describe how the data are compiled into a uniform and concise compilation in \S\ref{sec:compilation_pipeline}, which includes details on quality flag conversions, astrometric corrections, and the process for flagging sources that have multiple redshift measurements. The compilation also includes physical property measurements described in \S\ref{sec:SED} for a subset of galaxies with high-quality redshifts. In \S\ref{sec:results}, we describe the contents of the compilation, the spectroscopic completeness, stellar mass properties, types of star-forming and quiescent galaxies in the compilation, and the diverse range of environments included in the compilation. We showcase two application examples in \S\ref{sec:applications}: validating photometric redshifts with a detailed investigation of calibration metrics (\S\ref{sec:photoz_validation}) and using self-organizing maps (unsupervised machine learning) to gain insight into the galaxy sub-populations that do not currently have spectroscopic coverage (\S\ref{sec:SOMs}). In \S\ref{sec:future}, we discuss the future of the compilation, describing how the compilation is expected to grow as more datasets become publicly available and also provide details on how subsequent data releases will be carried out and made available for the community. 

Overall, the COSMOS Spectroscopic Redshift Compilation is intended to be a continuously evolving legacy resource for the community to be used for both calibration and scientific purposes. It can also be used to identify sources of interest that have multiple and sometimes contradicting redshifts from low S/N spectra (flagged within our compilation), thereby motivating deeper spectroscopic follow-up observations. {\textit{We highly encourage users of the compilation to not only cite this work but, if possible, also to cite the relevant references (provided in the compilation) associated with the original publication, particularly for cases where a small subsample of the compilation is being used.} The list of references are in the data release and will be updated as new data is included in the compilation. We also encourage the community to contact us\footnote{The community can contact \href{mailto:cosmosleadership@gmail.com}{cosmosleadership@gmail.com} for all queries and inclusion of new data sets in the compilation. Additionally, we highly encourage users to post in the ``issues'' and ``discussion'' sections of the \texttt{GitHub} repository \href{https://github.com/cosmosastro/speczcompilation}{\texttt{cosmosastro/speczcompilation}} in regards to queries and issues about the compilation and its usage.} and let us know about new data sets to be included in the compilation.

Throughout this paper, unless otherwise stated, we assume a $\Lambda$CDM cosmology ($H_0 = 70$ km s$^{-1}$ Mpc$^{-1}$, $\Omega_m = 0.3$, $\Omega_\Lambda = 0.7$), \cite{Chabrier2003} initial mass function for all stellar mass measurements, and the AB magnitude system \citep{OkeGunn1983}.

\section{Data}\label{sec:data}

There are a multitude of spectroscopic programs that observed targets within the COSMOS field from observer-frame optical to radio wavelengths. Several programs, such as $z$COSMOS \citep{Lilly2007}, GOGREEN \citep{Balogh2021}, and LEGA-C \citep{vanderWel2021} are publicly available through a centralized database (e.g., IRSA IPAC services, ESO Science Portal, Survey websites) allowing easy access. In many cases, program PIs opt to publish their spectroscopic redshifts and measurements within the VizieR catalog access tool \citep{Ochsenbein2000} that allows any user direct access to the full data set and also the ability to select objects of interest. However, the majority of spectroscopic redshifts and associated measurements can only be found within publications (e.g., survey papers, target-specific papers) and via private communication with program PIs.

Prior to 2018, the COSMOS Collaboration had a preliminary spectroscopic redshift compilation organized \& maintained by co-author MS where redshifts were collected via direct contact with PIs and via a systematic search of the NASA Astrophysical Database System (ADS) for papers with keywords such as ``COSMOS'', ``spectroscopic'', and also names of facilities (e.g., Keck, Gemini) and instruments (e.g., MOSFIRE, GMOS). We then updated this list for papers published after 2017. The search results were individually inspected for any reports of spectroscopic redshifts. 
In the majority of cases, redshifts are listed within a table(s) in the paper which we then manually gathered and organized into individual files. PI(s) and/or first authors were contacted in the case where coordinates were not made public or clarification on the data sets was needed. 

All data sets included in this first release of the compilation are described in Table \ref{table:datasets}. 

\begingroup 
\setlength\LTcapwidth{\textwidth} 
\setlength\LTleft{0pt}            
\setlength\LTright{0pt}           
\tiny
\setlength{\tabcolsep}{0.4pt}
\begin{longtable*}{@{\extracolsep{\fill}}llllccl}
    \caption{Description of all data sets included in the spectroscopic redshift compilation. The numerical ID number assigned to each survey and associated with the `survey' column in the compilation fits files are shown. Basic information regarding the telescope and instrument used for each data set is also shown, along with their respective redshift ranges. The type classification refers to the general wavelength range and are defined as the following: Opt (Optical; 0.35 -- 1 $\mu$m), NIR (near-infrared; 1 -- 5 $\mu$m), MIR (mid-infrared; 5 -- 40 $\mu$m), FIR (far-infrared; 40 -- 700 $\mu$m), and submm/mm ($>0.7$ mm). Programs are also classified if the telescope used is space- (S) or ground-based (G). The total number of objects, $N_\mathrm{obj}$, in each data set is also shown together with the appropriate reference(s). In the case where a publication is not available for a given dataset, we provide the name of the PI for reference.}
    \label{table:datasets}
      \cr \hline
      ID & Survey       & Telescope$/$Instrument & Type  & $z_\mathrm{spec}$   & $N_\mathrm{obj}$ & Reference(s) or PI \\
      \hline
      \endfirsthead
      \hline
      ID & Survey       & Telescope$/$Instrument & Type  & $z_\mathrm{spec}$   & $N_\mathrm{obj}$ & Reference(s) or PI \\
      \hline
      \endhead
      1     & 2dFGRS                    &  AAT/2dF                            & (G) Opt         &    0.00 -- 0.33   &   222     &  \citet{Colless2001} \\
      2     & 3DHST                     & {\it HST}/WFC3 \& \textit{HST}/ACS  & (S) Opt+NIR     &    0.00 -- 1.34   &   455     & \citet{Brammer2012}\\
            &                           &                                     &                 &                   &           & \citet{Momcheva2016}\\
      3     &   --                      & AAT                                 & (G) Opt         &    0.00 -- 3.29   &   636     & Suzuki et al. (\textit{in prep}) \\ 
      4     &   --                      & ALMA                                & (G) submm/mm    &    4.32 - 4.33    &   2       & \citet{Brinch2025}\\
      5     &   --                      & ALMA                                & (G) submm/mm    &    2.49 -- 2.51   &   7       & \citet{Champagne2021}\\
      6     &   --                      & ALMA                                & (G) submm/mm    &    2.78 -- 4.75   &   9       & \citet{Gentile2024}\\
      7     &   --                      & ALMA                                & (G) submm/mm    &   3.62 -- 5.85    &   4       & \citet{Jin2019} \\
      8     &   --                      & ALMA                                & (G) submm/mm    &   2.625           & 1         & \citet{Jin2024} \\
      9     &   --                      & ALMA                                & (G) submm/mm    &   6.75 -- 7.06    & 3         & \citet{Schouws2023} \\
     10     &    --                     & ALMA                                & (G) submm/mm    &   3.71 -- 3.72    & 2         & \citet{Schreiber2018_ALMA} \\
     11     &   --                      & ALMA                                & (G) submm/mm    &   4.820           & 1         & \citet{Sillassen2025} \\
     12     &   --                      & ALMA                                & (G) submm/mm    &   6.81 -- 6.85    & 2         & \citet{Smit2018} \\
     13     &   --                      & ALMA                                & (G) submm/mm    &   1.15 -- 1.63    & 55        & \citet{Valentino2018,Valentino2020a,Valentino2020b} \\
     14     & AS2COSPEC                 & ALMA                                & (G) submm/mm    &   1.90 -- 4.78    & 18        & \citet{Chen2022} \\
     15     & AzTEC COSMOS              & LMT/RSR \& SMA/SIS                  & (G) submm/mm    &   4.34            &  1        & \citet{Yun2015}  \\
     16     & AzTEC COSMOS              & ALMA \& NOEMA                       & (G) submm/mm    &  4.63             &  2        &  \citet{Jimenez2020} \\
     17     & BRiZELS	               & WHT/LIRIS \& TNG/NICS               & (G) NIR         &  0.84 -- 2.28     & 17        & \citet{Sobral2016} \\
     18     & C3R2-DR3                  & Keck/LRIS, DEIMOS, MOSFIRE          & (G) Opt+NIR     &  0.10 -- 4.17     & 347       & \citet{Stanford2021} \\
     19     & C3R2-LBT                  & LBT/LUCI1, LUCI2                    & (G) NIR         &  1.32 -- 2.56     & 163       & \citet{Saglia2022} \\
     20     & C3R2                      & Keck/LRIS, DEIMOS, MOSFIRE          & (G) Opt+NIR     & 0.00 -- 4.50      & 2869      & \citet{Masters2017,Masters2019}\\
     21     & CANDELSz7                 & VLT/FORS2                           & (G) Opt         & 0.66 -- 7.14      &   18      & \citet{Pentericci2018} \\
     22     & CAPERS                    & \textit{JWST}/NIRSpec               & (S) NIR         & 9.288             &  1        & \citet{Taylor2025} \\ 
     23     & CHESS                     & Keck/DEIMOS                         & (G) Opt         & 0.08 -- 1.61      &  233      & Lertprasertpong et al. (\textit{in prep}) \\
     24     & CHILES                    & ALMA                                & (G) submm/mm    & 0.01 - 1.29       &  80       & Blue Bird et al. \textit{in prep} \\ 
     25     & CLAMATO DR2               &  Keck/LRIS                          & (G) Opt         & 0.05 -- 3.52      &   600     & \citet{Lee2018} \\
            &                           &                                     &                 &                   &           & \citet{Horowitz2022}\\     
     26     & --                        & MMT/Binospec                        & (G) Opt         & 6.70 -- 6.88      & 9         & \citet{Endsley2022}\\ 
     27     & --                        & VLT/X-SHOOTER, Keck/DEIMOS          & (G) Opt         & 6.54 -- 6.60      & 2         & \citet{Sobral2015}\\ 
     28     & --                        & Keck/DEIMOS                         & (G) Opt         & 0.37 -- 6.32      & 22        & \citet{Brinch2024}\\
     29     & --                        & Keck/DEIMOS                         & (G) Opt         & 0.03 -- 6.47      & 3158      & PI: Peter Capak \\
     30     & --                        & Keck/DEIMOS                         & (G) Opt         & 0.12 -- 1.52      & 156       & \citet{Casey2012} \\
     31     & 10K-DEIMOS                & Keck/DEIMOS                         & (G) Opt         & 0.00 -- 6.60      & 10718     & \citet{Hasinger2018} \\
     32     & --                        & Keck/DEIMOS                         & (G) Opt         & 5.67 -- 5.75      & 6         & \citet{Henry2012} \\ 
      33    & --                        & Keck/DEIMOS                         & (G) Opt         & 0.00 -- 5.04      & 1059      & \citet{Kartaltepe2010} \\
      34    & --                        & Keck/DEIMOS                         & (G) Opt         & 0.39 -- 5.01      & 143       & PI: Mara Salvato \\ 
      35    & --                        & Keck/DEIMOS                         & (G) Opt         & 0.08 -- 1.76      &   225     & \citet{Shah2020} \\
      36    & --                        & Keck/DEIMOS                         & (G) Opt         & 6.54 -- 6.60      & 3         & \citet{Taylor2025_Keck}\\
      37    & DESI DR1                  & Mayall/DESI                         & (G) Opt         & 0.00 -- 6.81      & 96255     & \citet{AbdulKarim2025} \\
      38    & DESI DR2                  & Mayall/DESI                         & (G) Opt         & 0.00 -- 6.65      & 201445    & \citet{Ratajczak2025}\\
      39    & DESI-EDR                  &  Mayall/DESI                        & (G) Opt         & 0.00 -- 5.81      & 55327     & \citet{Adame2024} \\
      40    & FENIKS                    &  Keck/MOSFIRE                       & (G) NIR         & 3.34 -- 4.67      &    4      & \citet{AntwiDanso2025} \\
      41    & FMOS--COSMOS              & Subaru/FMOS                         & (G) NIR         & 0.37 -- 4.64      & 988       & \citet{Kartaltepe2015} \\ 
      42    & FMOS--COSMOS              & Subaru/FMOS                         & (G) NIR         & 0.70 -- 2.59      & 5484      & \citet{Kashino2019} \\
      43    & --                        & Subaru/FMOS                         & (G) NIR         & 0.60 -- 1.45      & 539       & PI: Tohru Nagao \\ 
      44    & --                        & Subaru/FMOS                         & (G) NIR         & 0.68 -- 1.57      &  85       & \citet{Roseboom2012} \\
      45    & --                        & Subaru/FOCAS                        & (G) Opt         & 0.05 -- 3.88      & 114       & PI: Tohru Nagao \\ 
      46    & --                        & Subaru/FOCAS                        & (G) Opt         & 0.828             & 1         & \citet{Zhu2025}\\
      47    & --                        & VLT/FORS2                           & (G) Opt         & 0.08 -- 5.49      & 1767      & \citet{Comparat2015} \\
      48    & --                        & VLT/FORS2                           & (G) Opt         & 2.44 -- 2.45      &   11      & \citet{Diener2015} \\
      49    & --                        & VLT/FORS2                           & (G) Opt         & 0.04 -- 1.29      & 530       & \citet{George2011} \\
      50    & --                        & VLT/FORS1                           & (G) Opt         & 0.03 -- 1.15      &   54      & \citet{Faure2011} \\      
      51    & GEEC2                     & Gemini-S/GMOS                       & (G) Opt         & 0.12 -- 1.51      &  807      & \citet{Balogh2011} \\
      52    & --                        & Gemini-S/GMOS                       & (G) Opt         & 0.10 -- 1.66      &  124      & Cox et al. (\textit{in prep}) \\
      53    & --                        & Gemini-S/GMOS                       & (G) Opt         & 0.8275            & 1         & \citet{Khostovan2025} \\
      54    & GOGREEN-DR1               & Gemini/GMOS                         & (G) Opt         & 0.50 -- 2.32      &  226      & \citet{Balogh2021} \\
      55    & HALO7D                    & Keck/DEIMOS                         & (G) Opt         & 0.29 -- 1.52      & 533       & \citet{Pharo2022}\\
      56    & --                        & MMT/Hectospec                       & (G) Opt         & 0.00 -- 1.44      &  277      & \citet{Sohn2019} \\
      57    & HETDEX-DR2                & HET/VIRUS                           & (G) Opt         & 0.01 -- 3.51      & 2447      & \citet{Mentuch2023} \\
      58    & HETVIPS                   & HET/VIRUS                           & (G) Opt         & 0.00 -- 2.10      & 325       & \citet{Zeimann2024}\\
      59    & HMS                       & Keck/LRIS, MOSFIRE                  & (G) Opt+NIR     & 1.36 -- 2.24      & 21        & \citet{Kriek2024} \\
      60    & HR-COSMOS                 & VLT/VIMOS                           & (G) Opt+NIR     & 0.74 -- 1.18      &   82      & \citet{Pelliccia2017} \\ 
      61    & \textit{HST}-Hyperion     & \textit{HST}/WFC3                   & (S) NIR         & 0.00 -- 5.99      & 12814     & \citet{Forrest2025}\\
      62    & --                        & Magellan/IMACS                      & (G) Opt         & 0.00 -- 4.66      & 2858      & \citet{Trump2007} \\
      63    & --                        & \it{Spitzer}/IRS                    & (S) MIR         & 0.59 -- 0.83      &   52      & \citet{Fu2010} \\
      64    & --                        & Keck/KCWI                           & (G) Opt         & 2.91 -- 2.92      & 3         & \citet{Daddi2021} \\
      65    & KMOS$^{3D}$               & VLT/KMOS                            & (G) NIR         & 0.38 -- 2.58      &  296      & \citet{Wisnioski2019}  \\
      66    & --                        & VLT/KMOS                            & (G) NIR         &  0.22 -- 0.90     &   10      & PI: Antonello Calabr\`o \\ 
      67    & KASH$z$                   & VLT/KMOS                            & (G) NIR         & 1.10 -- 1.63      &   15      & \citet{Harrison2016} \\
      68    & KROSS                     & VLT/KMOS                            & (G) NIR         & 2.49 -- 2.51      &   16      & \citet{Stott2016} \\
      69    & LAGER                     & Keck/LRIS                           & (G) Opt         & 6.90 -- 6.94      &   11      & \citet{Harish2022} \\
      70    & LAGER                     & Magellan/IMACS                      & (G) Opt         & 6.88 -- 6.94      & 6         & \citet{Hu2017}\\
      71    & LAGER                     & Magellan/IMACS \& LDSS3             & (G) Opt         & 6.90 -- 6.97      &   16      & \citet{Hu2021} \\
      72    & LEGA-C DR3                & VLT/VIMOS                           & (G) Opt         & 0.10 -- 4.31      & 3928      & \citet{vanderWel2021} \\
      73    & --                        & Keck/LRIS                           & (G) Opt         & 0.00 -- 4.54      &  447      & \citet{Casey2012} \\
      74    & --                        & Keck/LRIS                           & (G) Opt         & 2.27 -- 3.03      & 58        & \citet{Lee2016} \\  
      75    & --                        & LBT/LUCI1                           & (G) NIR         & 1.37 -- 2.22      & 11        & \citet{Maseda2014} \\
      76    & --                        & LBT/LUCI                            & (G) NIR         & 2.15 -- 2.20      & 31        & \citet{Polletta2021} \\
      77    & M2FS                      & Magellan/M2FS                       & (G) Opt         & 5.99 -- 6.15      & 3         & \citet{Fu2024} \\
      78    & M2FS                      & Magellan/M2FS                       & (G) Opt         & 5.64 -- 5.76      &   52      & \citet{Ning2020} \\
      79    & M2FS                      & Magellan/M2FS                       & (G) Opt         & 6.54 -- 6.63      & 7         & \citet{Ning2022} \\
      80    & MAGAZ3NE                  & Keck/MOSFIRE                        & (G) NIR         & 3.33 -- 3.43      &   22      & \citet{McConachie2022} \\  
      81    & --                        & Magellan/FIRE                       & (G) NIR         & 0.37 -- 0.95      &   25      & \citet{Calabro2018} \\
      82    & MAGIC                     & VLT/MUSE                            & (G) Opt         & 0.00 -- 6.61      &  2471     & \citet{Epinat2024}\\
      83    & --                        & ALMA                                & (G) submm/mm    & 5.85              &    2      & \citet{Casey2019} \\
      84    & --                        & MMT/Hectospec                       & (G) Opt         & 0.01 -- 2.29      &  291      & \citet{Prescott2006} \\
      85    & --                        & Magellan/IMACS                      & (G) Opt         & 0.19 -- 3.26      &   38      & \citet{Trump2009} \\
      86    & --                        & Subaru/MOIRCS                       & (G) NIR         & 1.43 -- 1.83      &   18      & \citet{Onodera2012} \\
      87    & --                        & Subaru/MOIRCS                       & (G) NIR         &  1.25 -- 2.09     &   34      & \citet{Onodera2015} \\
      88    & --                        & Subaru/MOIRCS                       & (G) NIR         &  2.19 -- 3.60     &   21      & \citet{Onodera2020} \\
      89    & --                        & Subaru/MOIRCS                       & (G) NIR         & 1.46 -- 1.63      &   12      & \citet{Tanaka2013} \\
      90    & MOSDEF                    & Keck/MOSFIRE                        & (G) NIR         & 0.80 -- 4.49      &   616     & \citet{Kriek2015} \\
      91    & --                        & Keck/MOSFIRE                        & (G) NIR         & 2.46 -- 2.49      &   42      & \citet{Casey2015} \\
      92    & --                        & Keck/MOSFIRE                        & (G) NIR         & 0.17 -- 3.58      &   32      & \citet{Casey2017} \\
      93    & --                        & Keck/MOSFIRE                        & (G) NIR         & 2.04 -- 2.99      &   52      & \citet{Darvish2020} \\
      94    & MAGAZ3NE                  & Keck/MOSFIRE                        & (G) NIR         & 2.06 -- 3.83      &   19      & \citet{Forrest2020,Forrest2022,Forrest2024}\\
            &                           &                                     &                 &                   &           & McConachie et al.\textit{in prep} \\
      95    & --                        & Keck/MOSFIRE                        & (G) NIR         & 2.76 -- 2.79      &    4      & \citet{Ito2023} \\
      96    & --                        & Keck/MOSFIRE                        & (G) NIR         & 4.53              &    1      & \citet{Kakimoto2024} \\
      97    & --                        & Keck/MOSFIRE                        & (G) NIR         & 2.97 -- 3.69      &   43      & \citet{Onodera2016} \\
      98    & --                        & Keck/MOSFIRE                        & (G) NIR         & 2.47 -- 3.78      &    6.     & \citet{Schreiber2018} \\
      99    & --                        & Keck/MOSFIRE                        & (G) NIR         & 0.76 -- 2.43      &   33      & PI: Nick Scoville \\
     100    & --                        & Keck/MOSFIRE                        & (G) NIR         & 2.14 -- 3.62      &  14       & \citet{Trakhtenbrot2016}\\
     101    & ZFIRE                     & Keck/MOSFIRE                        & (G) NIR         & 1.97 -- 2.31      &   90      & \citet{Tran2017} \\
     102    & --                        & Keck/MOSFIRE                        & (G) NIR         &  1.24 -- 3.42     &  151      & Vanderhoof et al. (\textit{in prep}) \\
     103    & SMUVS                     & VLT/MUSE                            & (G) Opt         &  0.07 -- 6.30     &  792      & \citet{Rosani2020} \\
     104    & MUSE GTO                  & VLT/MUSE                            & (G) Opt         &  1.50 -- 6.53     &  263      & \citet{Schmidt2021} \\         
     105    & --                        & VLT/MUSE                            & (G) Opt         &  0.36 -- 4.61     &   27      & \citet{Ventou2019} \\
     106    & --                        & \textit{JWST}/NIRSpec               & (S) NIR         & 14.44             &   1       & \citet{Naidu2025}\\ 
     107    & NICE                      & NOEMA                               & (G) submm/mm    & 1.64 -- 3.61      & 21        & \citet{Sillassen2024} \\
     108    & --                        & Keck/NIRSPEC                        & (G) NIR         & 2.03 -- 2.43      &    6      & PI: Jeyhan Kartaltepe\\
     109    & --                        & Keck/NIRSPEC                        & (G) NIR         & 3.14 -- 3.37      &    2      & \citet{Marsan2017} \\
     110    & --                        & \textit{JWST}/NIRSpec               & (S) NIR         & 7.00              &   1       & \citet{Akins2025} \\
     111    & --                        & \textit{JWST}/NIRSpec               & (S) NIR         & 7.04              &   1       & \citet{Akins2025b} \\
     112    & --                        & NOEMA, ALMA                         & (G) submm/mm    & 3.55 -- 4.10      &   5       & \citet{Jin2022} \\
     113    & --                        & NOEMA, Keck/MOSFIRE                 & (G) NIR+submm/mm& 5.10              & 1         & \citet{Shuntov2025_ER} \\
     114    & --                        & NOEMA, VLT/KMOS                     & (G) NIR+submm/mm& 2.49 -- 2.52      & 17        & \citet{Wang2016} \\
     115    & PRIMUS                    & Magellan/IMACS                      & (G) Opt         & 0.02 -- 4.60      & 29594     & \citet{Coil2011} \\
     116    & QUAIA                     & {\it Gaia}                          & (S) Opt         & 0.18 -- 4.12      &  289      & \citet{Storey2023} \\
     117    & REBELS                    & ALMA                                & (G) submm/mm    & 6.58 -- 7.68      &   11      & \citet{Bouwens2022} \\
     118    & SDSS DR16                 & SFT \& du Pont                      & (G) Opt         & 0.00 -- 6.58      & 1579      & \citet{Ahumada2020} \\
     119    & --                        & VLT/SINFONI                         & (G) NIR         & 0.97 -- 2.45      &   5       & \citet{Perna2015} \\
     120    & --                        & VLT/VIMOS                           & (G) Opt         & 0.69 -- 0.79      &  619      & \citet{Iovino2016}\\
     121    & --                        & VLT/VIMOS                           & (G) Opt 		   & 0.89 -- 2.09      & 34        & \citet{Gobat2017}\\
     122    & VIS$^{3}$COS              & VLT/VIMOS                           & (G) Opt         & 0.02 -- 3.59      &  696      & \citet{Paulino2018} \\
     123    & VUDS DR1                  & VLT/VIMOS                           & (G) Opt         & 0.04 -- 6.44      & 384       & \citet{LeFevre2015}\\ 
            &                           &                                     &                 &                   &           & \citet{Tasca2017} \\
     124    & VUDS DR2$^\star$     & VLT/VIMOS                           & (G) Opt         & 0.02 -- 6.44      & 4703      & Tasca et al. (\textit{in prep})\\ 
     125    & WERLS                     & Keck/MOSFIRE                        & (G) NIR         & 3.52 -- 6.93      &   83      & PIs: Caitlin Casey \& Jeyhan Kartaltepe \\
     126    & WERLS                     & Keck/MOSFIRE                        & (G) NIR         & 0.51 -- 8.23      &  192      & \citealt{Cooper2024} \\
     127    & --                        & {\it HST}/WFC3                      & (S) NIR         & 2.39 -- 3.23      &   10      & \citet{DEugenio2021} \\
            &                           &                                     &                 &                   &           & \citet{Lustig2021} \\     
     128    & --                        & {\it HST}/WFC3                      & (S) NIR         & 0.68 -- 2.18      &  150      & Based on 3D-\textit{HST} data \\ 
     129    & --                        & {\it HST}/WFC3                      & (S) NIR         & 1.84 -- 2.20      &   14      & \citet{Krogager2014} \\
     130    & XLS                       & VLT/X-SHOOTER                       & (G) Opt+NIR     & 2.07 -- 2.47      &   18      & \citet{Matthee2021} \\
     131    & {\it XMM}-COSMOS          & VLT/X-SHOOTER                       & (G) Opt+NIR     & 1.00 -- 1.60      &   10      & \citet{Brusa2015} \\     
     132    & CALYMHA                   & VLT/X-SHOOTER                       & (G) Opt+NIR     & 2.21 -- 2.23      &    3      & \citet{Sobral2018} \\
     133    & --                        & VLT/X-SHOOTER                       & (G) Opt+NIR     & 1.96 -- 2.69      &   14      & \citet{Stockmann2020} \\
     134    & --                        & VLT/X-SHOOTER                       & (G) Opt+NIR     & 3.78              &    1      & \citet{Valentino2020} \\
     135    & hCOSMOS                   & MMT/Hectospec                       & (G) Opt         & 0.00 -- 1.27      & 4399      & \citet{Damjanov2018} \\
     136    & zCOSMOS Bright            & VLT/VIMOS                           & (G) Opt         & 0.00 -- 4.45      & 20716     & \citet{Lilly2007,Lilly2009} \\
     137    & zCOSMOS Deep              & VLT/VIMOS                           & (G) Opt         & 0.00 -- 4.25      &  9371     & PI: Simon Lilly \\ 
     138    & zFIRE                     & Keck/MOSFIRE                        & (G) NIR         & 1.97 -- 3.53      &   216     & \citet{Nanayakkara2016}\\
    \hline
    \multicolumn{7}{l}{$^\star$\footnotesize Currently proprietary; however, this dataset will be made publicly available upon the VUDS DR2 official release.}
    \end{longtable*}
\endgroup

\phantomsection
\phantom{blah}

\section{Creating a Uniform Spectroscopic Compilation}
\label{sec:compilation_pipeline}

In this section, we describe in detail the various steps that are taken to generate our spectroscopic compilation including how quality flags and confidence levels are converted to our assumed system (\S\ref{sec:QF_scheme}), how astrometric corrections are applied to account for systematics in varying astrometry calibrations from program-to-program (\S\ref{sec:astrometry}), and how we flag objects that have multiple redshift measurements (\S\ref{sec:duplicates}). The compilation pipeline is set up so that it can easily incorporate new spectroscopic catalogs as they are made available, allowing for continuous updates.

\subsection{Quality Flags \& Confidence Levels}
\label{sec:QF_scheme}

Quality flag ($Q_f$) assessment schemes and their corresponding confidence levels (CL) vary from program to program. The most popular $Q_f$ scheme is the 0 -- 4 system in increments of 1 with some also including a $Q_f = 9$ in the case a single line is present. Some programs include an additional $10+X$ ($X = 0 - 4,9$) for sources that exhibit broad line features indicative of an AGN. Additional flags beyond the typical $Q_f$ described above are also included in other programs. For example, $z$COSMOS \citep{Lilly2007} includes a decimal flagging system ($X.1$, $X.3$, $X.4$, $X.5$; $X$ being the main flag class) to describe the spectroscopic-to-COSMOS2009 \citep{Ilbert2009} photometric redshift consistency. VUDS \citep{Tasca2017} has additional flags $X+20$, $+30$, $+40$, and $-10$ to highlight key features such as serendipitous objects, overlapping spectra, pair systems, etc. Other programs, such as DESI-EDR \citep{Adame2024}, use a completely different quality assessment flagging system that does not fall within the $Q_f$ 0 -- 4, 9 framework.

\begin{table}
    \caption{Adopted Quality Flags and Confidence Levels and their corresponding definitions. We add an additional factor of $10$ to the $Q_f$ in the case of broad line features where that information is available. For example, a $Q_f = 13$ would be indicative of a reliable redshift with broad emission line features.}
    \label{table:qfs}
    \begin{tabular*}{\columnwidth}{@{\extracolsep{\fill}}c c l}
    \hline
    $Q_f$ & CL & Description \\
          & (\%) &           \\
    \hline
    4 & 97 -- 100 & Very Reliable Redshift \\
    3 & 95        & Reliable Redshift \\
    2 & 80        & Moderate Detection of Emission Lines \\
    1 & 50        & Tentative Measurement \\
    0 & --        & No Measurement \\
    9 & 85        & Single Line/Ambiguous Detection with good S/N \\
    $+10$ &   --  & Indicative of Broad Line Feature \\
    \hline
    \end{tabular*}
\end{table}

Combining the numerous data sets into a single catalog requires normalizing the $Q_f$ and CL systems for all sources to ensure consistency. We adopt the following quality system as defined in Table \ref{table:qfs} and convert each program's quality flag to match our adopted system. For example, KMOS$^{3\rm{D}}$ \citep{Wisnioski2019} lists $Q_f = 1$ (redshift/detection is uncertain), $0$ (redshift is secure), and $-1$ (non-detection). We ignore all sources flagged as $-1$ and make the following conversions to our $Q_f$ system: $0 \rightarrow 4$ and $1 \rightarrow 1$. The MUSE GTO program \citep{Schmidt2021} uses a system ranging from 0 -- 3, where each flag is described as the following: Flag 0 -- difficult to ascertain if the emission line feature is real or not; Flag 1 -- a single trustworthy line detection of uncertain nature; Flag 2 -- classification with high confidence (multiple line detections $S/N < 5$ or unique line profiles; e.g., \lya); Flag 3 -- multiple line detections $S/N >5$ and no doubt about line classification. Our conversions are as follow $3 \rightarrow 4$, $2 \rightarrow 3$, $1 \rightarrow 9$, and $0 \rightarrow 1$. We refer to the associated paper, documentation, and/or direct communication with the author(s)/PI(s) to properly convert the original quality flags to our system. In the majority of cases, the conversion is straightforward as most programs follow a system similar to the $Q_f$ to that highlighted in Table \ref{table:qfs}. However, as seen in the subsection below, there are some cases where the conversion is not as straightforward.

\subsubsection{DESI Quality Flag Conversion}
We use the full DESI-EDR v1.0 catalog\footnote{\url{https://data.desi.lbl.gov/public/edr/vac/edr/zcat/fuji/v1.0/zall-tilecumulative-edr-vac.fits}} in our compilation and use \citep{Adame2024} as the main documentation for applying our $Q_f$ scheme. Spectroscopic redshifts were assigned using an automated approach (\texttt{Redrock}\footnote{\url{https://github.com/desihub/redrock/tree/0.15.4}}) that generates Principal Component Analysis (PCA) templates, fits them to every target, and identifies the best redshift and spectral type (galaxy, QSO, or star; labeled in \texttt{SPECTYPE}) based on the lowest $\chi^2$. \texttt{Redrock} also generates a bitmask flag, \texttt{ZWARN}, that identifies any key issues that may have occurred during spectroscopic redshift assignment (e.g., narrow wavelength coverage, $\chi^2$ of best fit is too close to the second best, fiber was unplugged/broken, no data in fiber). Limiting the catalog to \texttt{ZWARN = 0} removes sources that have known issues in their spectroscopic redshift determination. This also corresponds to \texttt{DELTACHI2} ($\Delta \chi^2$) $ > 9$, which identifies the separation between the first and second best model. Larger \texttt{DELTACHI2} corresponds to greater statistical confidence that the template that best fits the object and the assigned redshift is correct.

\citet{Adame2024} (see their \S4.2.2) suggests a scheme to select `good' redshifts with catastrophic failure fractions $<0.5\%$, which we consider as $Q_f = 4$. We refer to this criteria scheme to assign $Q_f$ to the DESI-EDR sources. We first start by limiting the samples to \texttt{ZWARN = 0} and $z > 0$ where the latter removes objects that are slightly negative in redshift and have \texttt{SPECTYPE = GALAXY} (issue with redshift determination). These sources were initially assigned a $Q_f = 2$ given that there were no known issues in the redshift determination; however, a subset of them do not have reliable $\Delta\chi^2$ to separate between the different spectral types. Sources satisfying the following criteria are then assigned a $Q_f = 4$:
\begin{itemize}
    \item[\ding{118}] Luminous Red Galaxy: $z > 0.4$ \& $z < 1.1$ \& $\Delta\chi^2 >15$ 
    \item[\ding{118}] Bright Galaxy: $z > 0.01$ \& $z < 0.6$ \& $\Delta\chi^2 > 40$
    \item[\ding{118}] \oii~Emitter: $z > 0.6$ \& $z < 1.6$ \& $\Delta\chi^2 > 8$
    \item[\ding{118}] \lya~Emitter: $z > 2$ \& $\Delta\chi^2 > 8$
    \item[\ding{118}] QSO: $z > 0.6$ \& $z < 3.5$ \& \texttt{SPECTYPE = QSO}
\end{itemize}
Note that these selection criteria are taken from the DESI-EDR documentation \citep{Adame2024} with direct consultation with the PI. For all sources, if \texttt{SPECTYPE = QSO}, then an additional factor of $+10$ is also assigned to the final $Q_f$. The DESI DR1 \citep{AbdulKarim2025} and DESI DR2/COSMOS \citep{Ratajczak2025} were also converted by using associated documentation citing conditions resulting in minimal catastrophic errors and were also validated with direct communication with the associated PI.

\subsection{Astrometric Corrections and COSMOS Catalogs}
\label{sec:astrometry}
The programs included in this compilation have different astrometric calibrations such that when combined into a single catalog, can result in variable systematic astrometric uncertainties. Therefore, we require that the final compilation has astrometric corrections that bring all programs to the same calibration. 

We start by first matching each program to the COSMOS2020 Classic catalog \citep{Weaver2022} with a $1.5''$ search radius which is astrometrically calibrated with GAIA DR1 data. A total of 263823 of 274924 sources ($\sim96\%$) matched and were used to measure the astrometric offset. This was done by calculating the difference in RA and Dec.\ for each source with reference to the COSMOS2020 Classic RA and Dec. The median RA and Dec.\ offset per program was then applied to the respective sources. The typical correction applied was $\sim88$ mas with a median absolute deviation of $\sim48$ mas. The applied astrometric corrections are checked by measuring the offsets per program between the corrected RA and Dec.\ and the coordinates published in the COSMOS2020 Classic catalog, but this time with a $1''$ search radius. The typical offset per program after correction was $\sim69.6$ mas with a median absolute deviation of $\sim35$ mas. Figure \ref{fig:astrometry_check} shows the astrometric offsets for all sources in the compilation with a COSMOS2020 Classic match after the corrections described above were applied.

After applying these corrections, we match the spectroscopic compilation with both versions of the COSMOS2020 catalog (Classic and Farmer; \citealt{Weaver2022}) and COSMOS2009 \citep{Ilbert2009} catalogs with a $1''$ search radius. The RA and Dec.\ of each source in the ancillary COSMOS catalogs are included in the compilation (see Table \ref{table:comp}). Photo-$z$ information is also included for each source using \lephare~measurements in COSMOS2020 Classic.
 
\begin{figure}
    \centering
    \includegraphics[width=\columnwidth]{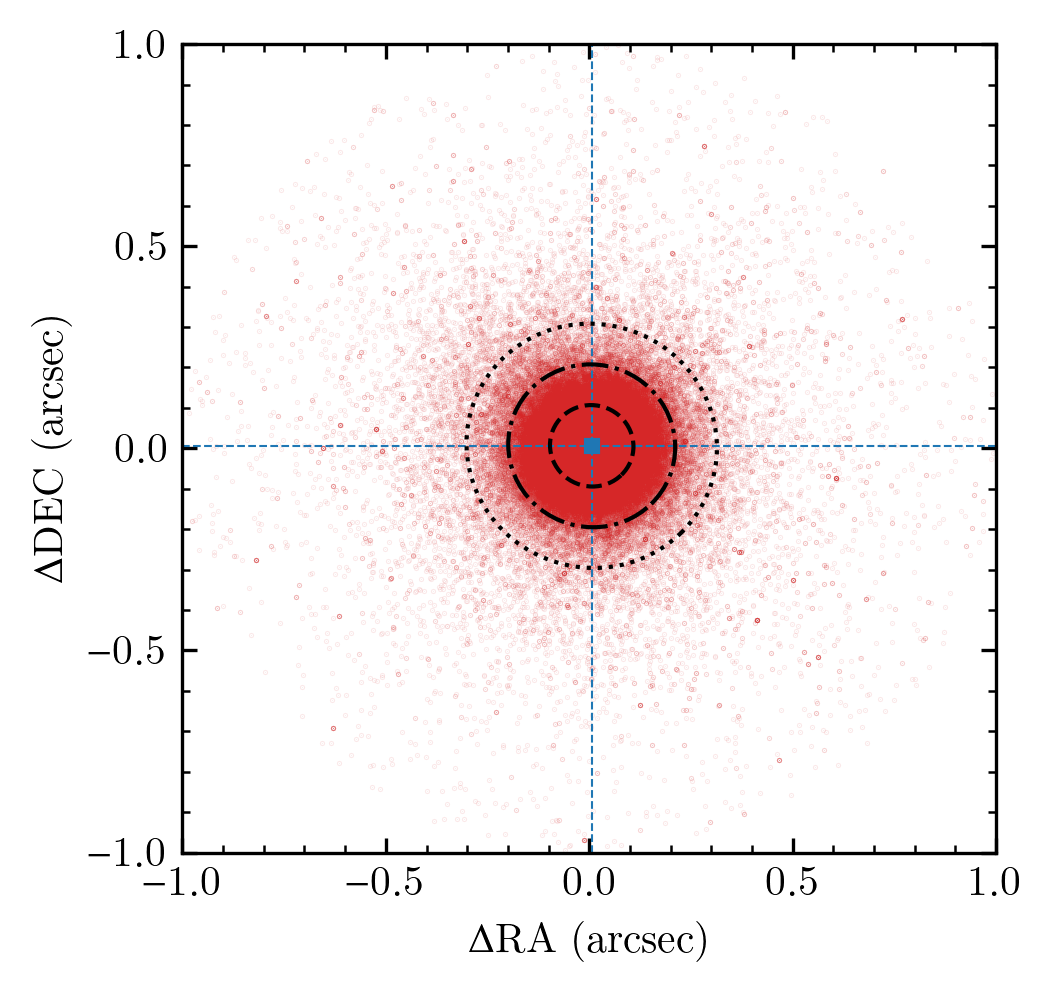}
    \caption{Offsets in the astrometry of all sources in the compilation after astrometric corrections were applied using COSMOS2020 Classic as the reference catalog. The \textit{dashed}, \textit{dash-dotted}, and \textit{dotted} lines are the $1\sigma$, $2\sigma$, and $3\sigma$ confidence levels. The median offset is shown as the \textit{blue square} and  is $\sim69.6$ mas with a median absolute deviation of $\sim 35$ mas and standard deviation of $105$ mas.}
    \label{fig:astrometry_check}
\end{figure}

\subsection{Mitigating Pair Galaxy Effects}

One issue that was noted in matching the compilation to COSMOS2020 Classic was the situation where line-of-sight (LoS) pairs were matched to a single COSMOS2020 galaxy (e.g., only one source detected). As a result, a single COSMOS2020 ID could potentially have wildly different redshifts. To mitigate this effect, we rely on the COSMOS ACS/F814W catalog \citep{Leauthaud2007} given its higher resolution compared to the ground-based COSMOS2020 catalog to resolve the pairs. Within the COSMOS2020 catalog are associated IDs (\texttt{ID\_ACS}) that connect to the ACS catalog. As such, we cross-match both the ACS and COSMOS2020 catalogs with the compilation. We then check to see for a given source in the compilation if the COSMOS2020 \texttt{ID\_ACS} and the ACS catalog ID equate. If not, then that specific match is flagged out by replacing the \texttt{Id\_COS20\_Classic} in the compilation as $-999$ (no Classic counterpart). This step is done in both determining (1.5$''$ search radius) and applying (1$''$ search radius) astrometric corrections to mitigate the incorrect matching effects. 

We note that this is not a fully guaranteed solution to this problem where some sources, especially at high-$z$, may lack an ACS/F814W detection. As such, we do caution users to double check for LoS pair galaxies especially in the case where redshifts differ significantly and/or expected redshifts are not met. Future higher resolution imaging can help alleviate this issue.

\subsection{Flagging Duplicate Sources}
\label{sec:duplicates}
Duplicate sources exist in the main compilation when two or more programs have observed the same source under different spectroscopic configurations. These sources can be of great interest for several reasons. For example, each independent program may have reported spectroscopic redshifts that are in agreement giving more confidence that the spectroscopic redshift is accurate. There may also be cases where spectroscopic redshifts are not in agreement and that tension, especially if $Q_f = 4$ for each reported redshift, could warrant further investigation. Duplicate sources in the compilation could also be useful to gauge what spectroscopic configurations have been used so far to observe an individual source (e.g., wavelength coverage, spectral resolution) for which the user could contact the PIs of the program requesting the 1D and 2D spectra enabling wider wavelength coverage for spectroscopic analyses. 

We identified duplicate sources by using the matched COSMOS2020 Classic, Farmer, COSMOS2015, and COSMOS2009 catalogs to identify duplicate IDs in each respective catalog that ended up in our main spectroscopic compilation. This was done by first separating the compilation to investigate only those galaxies that matched the COSMOS2020 Classic. An internal match is done for these sources based on the COSMOS2020 Classic ID where duplicate sources in our compilation would have the same Classic ID. For each source that has the same ID, we test which one has a better $Q_f$ and assign that specific compilation source with a priority flag of 1 (see Table \ref{table:comp}). We also assign a priority flag of 1 to the most recent redshift measurement in the case that two or more spectroscopic redshifts per duplicate source have the same $Q_f$. 

After inspecting all COSMOS2020 Classic matches, we then repeat the same procedure but for the sources with a Farmer match but no Classic match and repeat the above procedure. This is followed by COSMOS2015 and then COSMOS2009 sources. For the sources that do not match in any of the COSMOS catalogs, we flag duplicate sources by doing an internal coordinate match of these subsets within a $1''$ search radius and apply the same flagging method as described above. In total, there are 37,510 unique groups corresponding to 104,681 redshifts and a total of \uniqueSources~sources.

\section{Constraining Physical Properties}
\label{sec:SED}


Although sources in the published COSMOS catalogs include spectral energy distribution (SED) fits and derived physical properties, we opted to reassess these SED fits by fixing the redshift to the spectroscopic redshift. This provides a more reliable measurement of the physical properties associated with each galaxy, especially for those where the photometric redshifts are significantly different from the spectroscopic redshift. In this section, we describe how we constrain the physical properties of galaxies within the compilation using two widely used SED fitting codes: \cigale~\citep{Boquien2019,Yang2020} and \lephare~\citep{Arnouts1999,Ilbert2006}. Multi-wavelength photometry from COSMOS2020 Classic is used when available to constrain the SEDs for each unique source with the redshift fixed to the measured spectroscopic redshift. This corresponds to 133,916~unique sources (92\% of all unique sources within COSMOS2020 coverage) with COSMOS2020 Classic photometry ranging from CFHT/MegaCam \textit{u} to \textit{Spitzer}/IRAC 4.5$\mu$m that are used for SED fitting and are provided to the community within the data release. However, results shown below are restricted to SED fits for sources that have $Q_f = 3$ and $4$, resulting in a total of 82,261 COSMOS2020 Classic-matched $0 < z < 7.7$ unique sources. This ensures that sources identified as BL-AGNs are not included in the results; however, not all programs explicitly labeled broad line features and the sample includes X-ray and IR AGN. As such, we caution users when interpreting the SED-derived physical properties for sources that are AGN in the subset and encourage users to apply AGN-specific templates and models in the SED fitting process. 

\subsection{\cigale}

We use \cigale~\citep{Boquien2019}, a grid-based template SED fitting suite, to derive physical properties and constrain SEDs using the best spec-$z$ for each source. The setup files associated with our \cigale~SED fits are included in the data release and we briefly describe them here. We assume a \cite{Chabrier2003} initial mass function (IMF) with the \cite{BC2003} synthetic stellar population model and stellar metallicities in the range of 0.0004 to 0.05 (0.02 -- 2.5 $Z_\odot$). A delayed-$\tau$ star formation history is assumed and contains two components: a main/old stellar population and a young population formed in a recent burst. We define the age of the main population in the range of 50 Myr to 13 Gyr with an $e$-folding time scale ($\tau_\textrm{main}$) between 50 Myr and 20 Gyr. The ages of the young stellar population formed in a recent burst are evaluated between 5 and 50 Myr with $\tau_\textrm{burst}$ between 5 to 100 Myr and a fractional contribution to total stellar mass considered between 0 and 90\%, where 0\% would imply no recent burst.

Nebular emission line contributions are also included and are based on the \cite{Inoue2011} templates generated using \verb|Cloudy 13.01| \citep{Ferland1998,Ferland2013} with a range of ionization parameters, $\log_{10} U$, between $-4$ and $-1$, gas-phase metallicities between 0.0004 and 0.051 (0.02 -- 2.5 $Z_\odot$), fixed electron density of 100 cm$^{-3}$, LyC escape fractions of 0\% with no fraction absorbed by dust, and fixed emission line widths of 300 km s$^{-1}$. We assume a \cite{Calzetti2000} dust attenuation curve with no 2175\,\AA~bump and $E(B-V) \sim 0 - 2$ mag and an LMC dust extinction curve \cite{Pei1992} for the emission lines. Lastly, we apply the \cite{Draine2014} dust emission model assuming a constant PAH mass fraction of 2.5\%, minimum radiation field $U_\textrm{min} = 1$ with $\alpha = 2$ defining the slope of the radiation field ($\textrm{d}U/\textrm{d}M \propto U^\alpha$), and $\gamma = 0.1$ (fraction of starlight illuminated from dust). 

\subsection{\lephare}

Physical properties such as absolute magnitudes, stellar masses and SFR are also estimated using the template-fitting code \lephare~\citep{Arnouts1999,Ilbert2006} with the same configuration as used in COSMOS2020 \citep{Weaver2022}, except that the redshift is now fixed to the spectroscopic redshift reported in our compilation.

\begin{figure*}
    \centering
    \includegraphics[width=\textwidth]{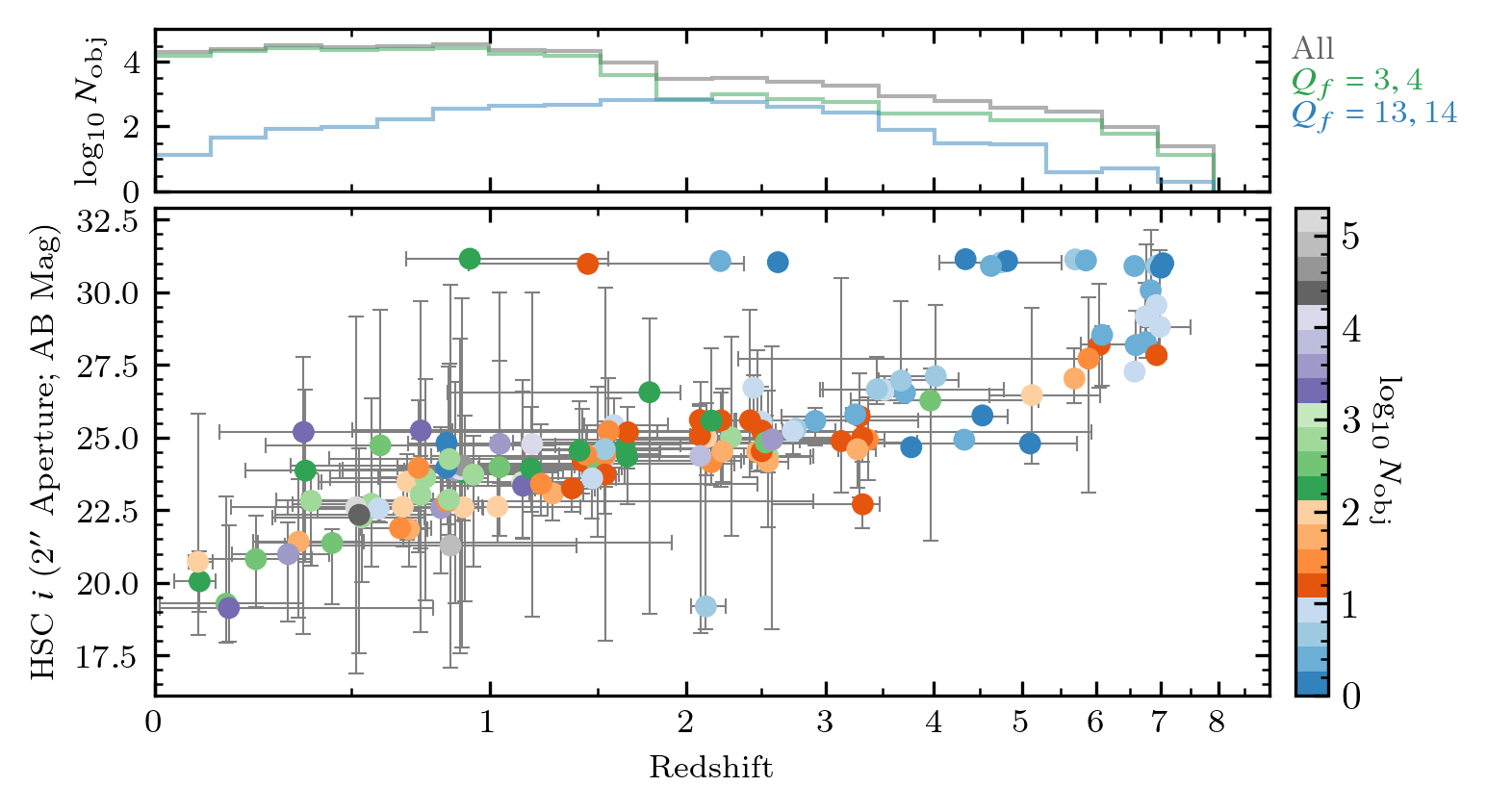}
    \caption{\textit{Top Panel} shows the published spectroscopic redshift distribution of all sources within the final compilation (\textit{grey}) along with those that have high-quality spec-$z$ (\qf$ = 3 - 4$) and broad-line features (\qf$ = 13 - 14$). The range of HSC $i$ magnitudes and spec-$z$ for each individual program is shown in the \textit{lower panel} with each symbol corresponding to the median $i$ magnitude and redshift for each program. Error bars represent the 16$^\textrm{th}$ and 84$^\textrm{th}$ percentile ranges. Programs with no COSMOS2020 Classic matches are placed at 31 mag with a slight perturbation for visualization purposes. In total, \uniqueSources~sources comprise the total compilation with \allSources~redshift measurements from \nprograms~programs.}
    \label{fig:zdistrib}
\end{figure*}

The model assumptions are described in \citet{Ilbert2015}. We built our set of templates using the \cite{BC2003} synthetic stellar population model and assume a \cite{Chabrier2003} IMF. We assume six different star formation histories (SFH): four exponentially declining ($\tau =  1$, 3, 5, 30 Gyr) and two delayed-$\tau$ (with $\tau = 1$, 3 Gyr). We assume two stellar metallicities (0.5$Z_\odot$ and $Z_\odot$) and 43 ages ranging from 0.05 Gyr to the maximum age of the Universe at the considered redshift. We add emission lines to the \cite{BC2003} templates by adopting the empirical relation from \citet{Ilbert2009} that links the UV intrinsic luminosity and the \oii~flux. We assume constant ratios between emission lines prior to applying dust attenuation. The emission line fluxes are allowed to vary all-together by 50\% around the fiducial value during the fit. We include two different attenuation curves \citep{Calzetti2000, Arnouts2013} and consider $E(B-V) = 0 - 0.7$ mag.  The intergalactic medium (IGM) absorption is accounted for by using the analytical correction of \citet{Madau1995}.
 
While \lephare{} outputs several estimates of the physical parameters, we select the stellar masses and the SFR from the template which minimizes the $\chi^2$.

\section{The COSMOS Spec-$z$ Compilation}
\label{sec:results}

The compilation consists of two key files: \texttt{\_all.fits}  (all redshift measurements) and \texttt{\_unique.fits} (only unique sources; see \S\ref{sec:duplicates}). A total of \uniqueSources~unique sources make up the compilation with \uniqueReliableSources~and \uniqueBLSources~sources having \qf~$= 3$ -- 4 and \qf~$= 13$ -- 14, respectively. The full compilation consists of \allSources~individual redshift measurements, where \allreliableSources~and \allreliableBLSources~redshifts have \qf~$= 3$ -- 4 and \qf~$=13$ -- 14, respectively. Sources with multiple measurements (groups) are also identified within the full compilation, where we find \NGroups~groups consisting of \AvgGroupSize~members on average with some groups having up to \MaxGroupSize~redshift measurements. 

Table \ref{table:datasets} highlights all \nprograms~programs that comprise the compilation within a 10 deg$^2$ area. The wider area compared to the 2 deg$^2$ coverage of COSMOS is in preparation of next-generation wide-area surveys. In total, \nprogramsGROUND~and \nprogramsSPACE~are based on ground- and space-based telescopes/observatories, respectively. The number of space-based redshifts is expected to increase as more data becomes available via missions such as \textit{JWST}, \textit{Euclid}, \textit{Roman}, and \textit{SPHEREx}. The majority of redshifts are drawn from \nprogramsOPT~optical ($0.35 - 1$ $\mu$m)~programs and  \nprogramsNIR~near-infrared ($1 - 5$ $\mu$m) programs. There are also \nprogramsRADIO~millimeter/submillimeter ($>0.7$ mm) programs primarily drawn from ALMA and NOEMA. Only \nprogramsMIR~mid-infrared ($5 - 40$ $\mu$m) program is currently in the compilation; however, that is expected to increase as \textit{JWST}/MIRI spectroscopic programs publish new redshifts.

The top panel of Figure \ref{fig:zdistrib} highlights the redshift distribution of the full compilation and the two high \qf~subsets. Both the full compilation and  \qf~$=3 - 4$ subset have a median redshift of $z\sim0.7$, while the broad line subset (\qf~$=13$ -- 14) has a median redshift of $z \sim 1.8$, where the latter is most likely due to programs specifically targeting high-$z$ AGN and quasars. Although the compilation contains primarily low-$z$ ($z<1$) galaxies, we find that $\sim10$\% (7\%) of the full (\qf~$=3 - 4$) compilation contains $z > 2$ galaxies, which is due to the relatively recent increase in availability of near-IR spectroscopy (enabling the detection of \halpha, \oiii, \hbeta, \oii~emission) and \lya~emission follow-up via optical spectroscopy. Given that \oiii~emission can be observed from the ground up to $z \sim 3.5$ and the faint nature of \oii~emission with increasing redshift (e.g., \citealt{Khostovan2015,Khostovan2016,Khostovan2020}),  measurements in the compilation with $z > 3 - 4$ are observed primarily based on \lya~emission (up to $z \sim 7$ within optical spectroscopy), as well as other rest-frame UV lines (e.g., C {\sc iv}). 

The main panel of Figure \ref{fig:zdistrib} highlights the range of Subaru/HSC~$i$ total magnitudes based on $2''$ apertures drawn from COSMOS2020/Classic. Each data point refers to a single survey as listed in Table \ref{table:datasets} with error bars showing the redshift range and $i$ magnitude in the survey. The majority of datasets forming the compilation are found to have $i$ magnitudes of 22 -- 25 mag at $0.5 < z < 2.5$ and $i > 25$ mag at $z > 2.5$, highlighting the observationally faint nature of high-$z$ sources (although intrinsically bright; e.g., Malmquist bias). We note that some of the surveys did not have COSMOS2020/Classic matches and are artificially assigned $i = 31$ mag simply to represent the datasets in Figure \ref{fig:zdistrib}. Interestingly, the $z < 0.5$ data sets included in the compilation are primarily biased towards $i < 22.5$ mag and do not contain significantly faint sources.

\begin{figure}
	\centering
	\includegraphics[width=\columnwidth]{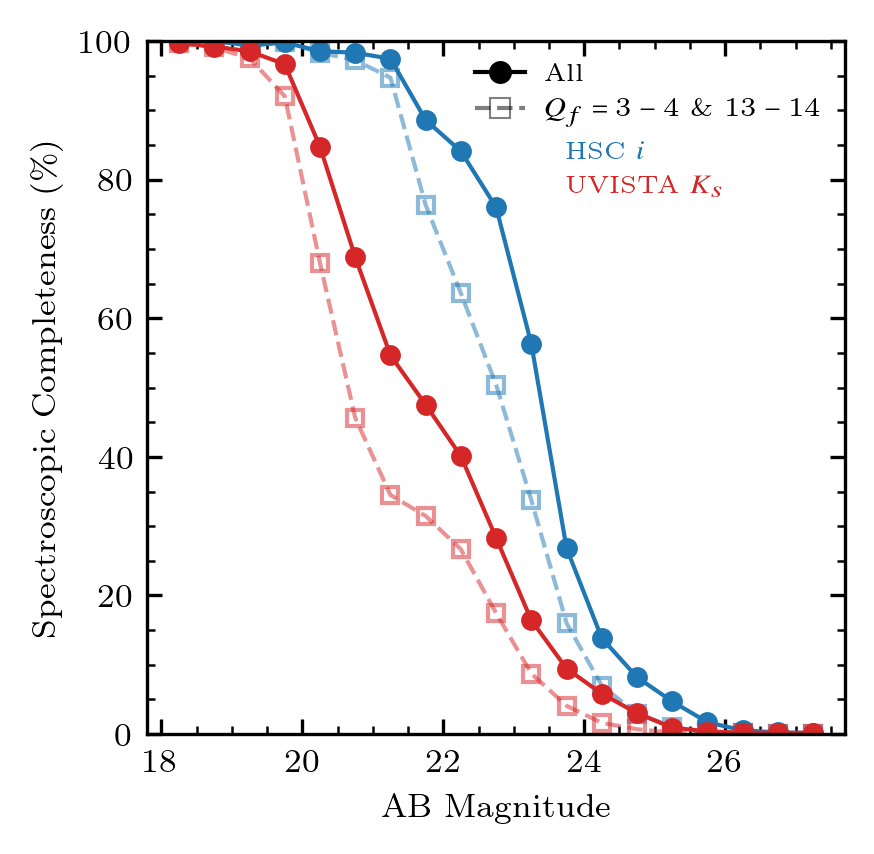}
	\caption{The spectroscopic completeness in terms of HSC $i$ (\textit{blue}) and UltraVISTA $K_s$ (\textit{red}) magnitudes with the full population and high \qf~subset shown as \textit{solid circles} and \textit{empty squares}, respectively. We find 50\% spectroscopic completeness for the full (high \qf) sample at $i \sim 23.4$ ($22.8$) mag and $K_s \sim 21.6$ ($\sim 20.7$) mag. Note this does not include any redshift dependency.}
	\label{fig:completeness}
\end{figure}

\subsection{Completeness of the Compilation}
\label{sec:completeness}

\begin{figure*}
	\centering
	\includegraphics[width=\textwidth]{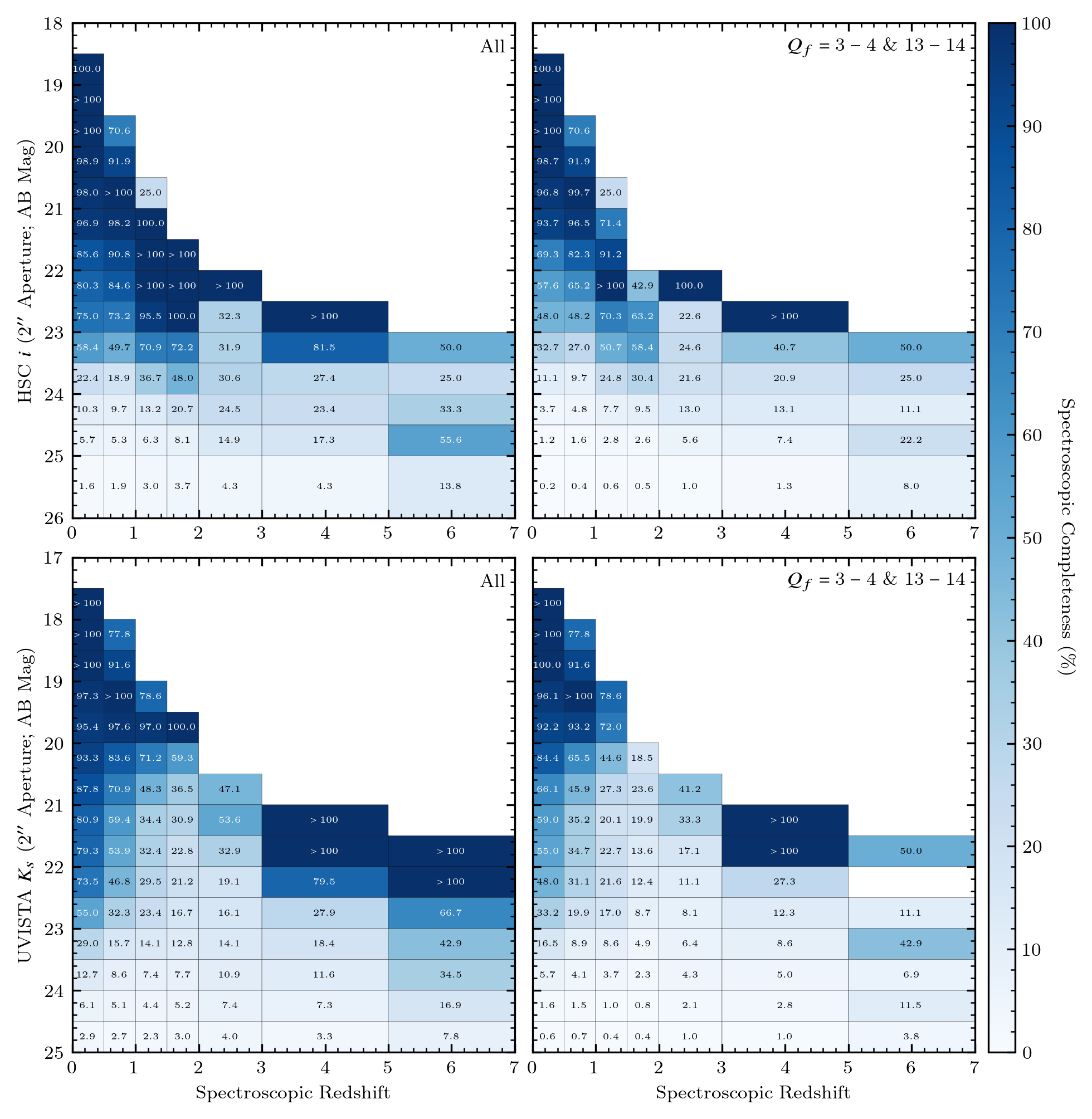}
	\caption{The spectroscopic completeness (in percentage) based on HSC $i$ (\textit{top panel}) and UltraVISTA $K_s$ (\textit{bottom panel}) magnitudes factoring in redshift dependency. The \textit{left} panels are for the full compilation and the \textit{right} panels are for the high \qf~subset. The 50\% completeness limit is roughly constant at $i \sim 23 - 23.5$ mag but is variable in the high \qf~subset shifting to fainter magnitudes from $z \sim 0$ to $z \sim 2$. The 50\% completeness limit in $K_s$ is variable in both the full compilation and high \qf~subset. Note that $>100\%$ is expected and signifies disagreements between COSMOS2020 photometric redshifts and the spectroscopic redshifts in our compilation. Overall, this highlights the need to understand the completeness not just of the whole sample, but also at various redshift and magnitude slices.}
	\label{fig:completeness_2D}
\end{figure*}

\begin{figure*}
	\centering
	\includegraphics[width=\textwidth]{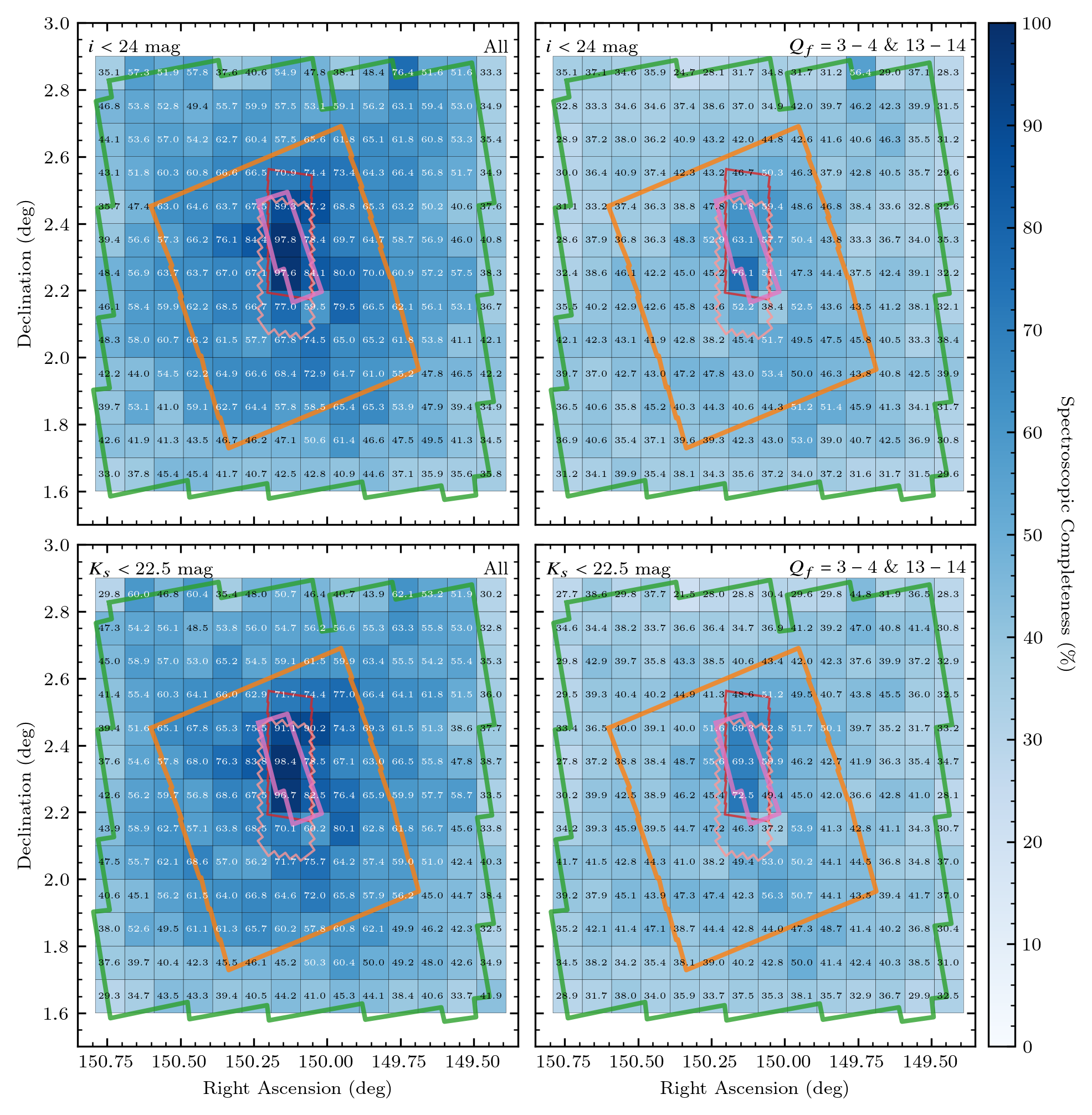}
	\caption{The spatial spectroscopic completeness of the compilation limited to $i < 24$ mag (\textit{top}) and $K_s < 22.5$ mag (\textit{bottom}) with the full (high \qf) samples shown on the \textit{left} (\textit{right}) panels. Each numerical value corresponds to the spectroscopic completeness of that specific 0.01 deg$^2$ region. The central region has the highest spectroscopic completeness due to the many follow-up programs focused on the CANDELS \textit{HST}/ACS (\textit{light red}) and WFC3 (\textit{red}) region (e.g., 3D-\textit{HST} -- \citealt{Brammer2012}; MOSDEF -- \citealt{Kriek2015}; HETDEX -- \citealt{Mentuch2023}). The outer regions also have $\gtrsim 30$\% completeness even for the high \qf~subset reaching the edges of the \textit{HST}/ACS F814W imaging (\textit{green}). Future \textit{JWST} spectroscopic follow-up within COSMOS-Web (\textit{orange}) and PRIMER (\textit{pink}) and other COSMOS-based programs including PASSAGE (PI: Matthew Malkan) and COSMOS-3D (PI: Koki Kakiichi) will both increase the spatial spectroscopic completeness as well as extend it to fainter magnitudes.}
	\label{fig:completeness_spatial}
\end{figure*}

Measuring the selection completeness function of the compilation is quite difficult given the diverse range of selection functions and observational biases that encompass the full data set. However, we can address the spectroscopic completeness of the compilation, which we define as the fraction of galaxies within a galaxy property range that has spectroscopic confirmation compared to the total population of galaxies observed within the same associated range. 

We use the COSMOS2020/Classic catalog as our basis for the total galaxy population and limit our compilation to only those with matches (93.5\% of all sources within the 2 deg$^2$ COSMOS coverage in the compilation). Figure \ref{fig:completeness} shows the completeness fraction in terms of HSC $i$ and UltraVISTA $K_s$ magnitudes for both ``all'' sources with spectroscopic redshifts and those restricted to a high \qf. We find that the full sample is 50\% complete down to $23.4$ ($i$) and $21.6$ ($K_s$) mag. Restricting this to just the higher \qf~flags, we find a 50\% completeness limit of $22.8$ ($i$) and $20.7$ ($K_s$) mag. This does not take into account the redshift distribution and, as shown in Figure \ref{fig:zdistrib}, the spectroscopic compilation contains numerous $z < 1$ galaxies that typically have $i < 25$ mag, while $z > 1$ sources are, as expected, biased towards fainter $i$ magnitudes.

Figure \ref{fig:completeness_2D} shows the completeness limits as a function of redshift and magnitude. For each grid point, we measure the fraction of galaxies that have similar magnitudes and redshifts in the compilation compared to the full population of galaxies. Given that COSMOS2020 only has photometric redshifts and that there will be disagreements between photometric and spectroscopic redshifts, it is therefore expected that some of the grid points will have $>100$\% completeness. 

We find that $z < 1$ galaxies (with all \qf) show 50\% completeness at $i < 23.5$ ($K_s < 22.5 - 23$) mag consistent with the case where redshift distributions are not factored in the completeness measurement (see Figure \ref{fig:completeness}). Limiting the samples to high \qf, we find a 50\% completeness limit of $i < 22.5 - 22.5$ mag and $K_S < 20.5 - 22$ mag for $z < 1$ galaxies. At $1 <  z < 2$, we find that 50\% completeness to be consistent with $i \sim 23.5 - 24$ mag and $K_s \sim 20.5 - 21$ for the full compilation and $i \sim 23.5$ mag and $K_s \sim 20$ mag when limited to high \qf. 

The spectroscopic completeness at $z > 2$ for the full sample is mostly consistent with $i \sim 22.5 - 23.5$ at the 50\% level. In terms of $K_s$ magnitudes, we find the 50\% level reaches $\sim 23$ mag at $5 < z < 7$ for the full sample and is driven by spectroscopic follow-up of \lya~emitters in the observer-frame optical. Such sources typically have blue UV spectral slopes with the $K_s$ band corresponding to the rest-frame $\sim 2500$ -- $4000$\AA. However, limiting to high \qf~essentially removes this $> 50\%$ completeness primarily due to the difficulty of reliably confirming high-$z$ targets. We note a spectroscopic completeness $> 50\%$ for high \qf~samples with $K_s < 22$ mag at $3 < z < 4$, which is based on confirmation of \oiii+\hbeta~emitters in the near-IR as well as \lya~emitters in the observer-frame optical.

Lastly, we investigate the spatial completeness by calculating the fraction of galaxies within the compilation relative to the COSMOS2020/Classic catalog within a given area in COSMOS. Figure \ref{fig:completeness_spatial} shows the spatial completeness for the full 2 deg$^2$ field limited to galaxies with $i < 24$ and $K_s < 22.5$ mag. We apply this restriction as the full spectroscopic completeness decreases significantly at fainter magnitudes, as shown in Figure \ref{fig:completeness}. Not applying these faint cutoffs would result in low spatial completeness measurements solely due to the overall low spectroscopic completeness at fainter magnitudes.

\begin{figure}
	\centering
	\includegraphics[width=\columnwidth]{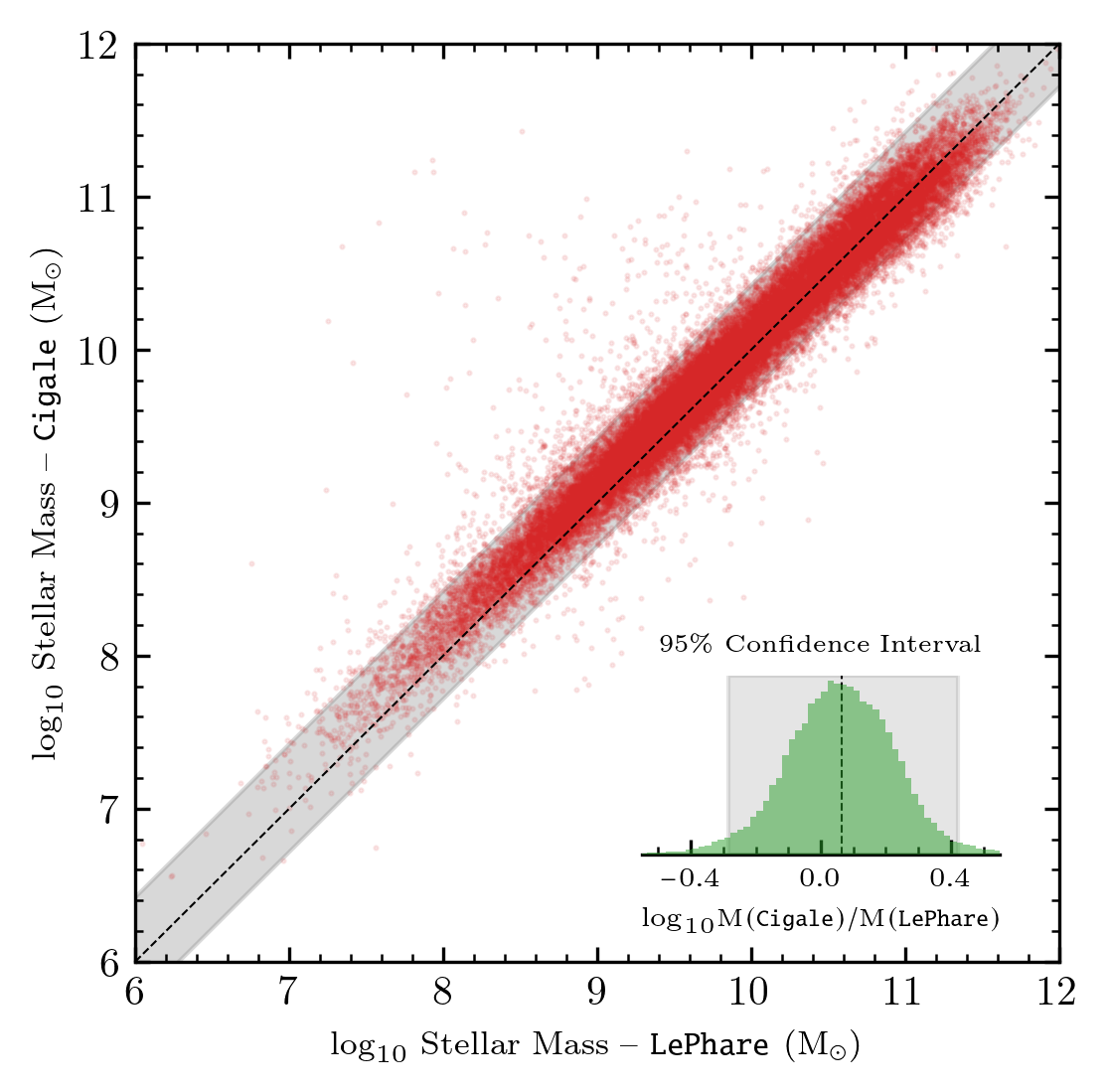}
	\caption{Comparison of stellar mass measurements from \cigale~and \lephare~for all sources with \qf$=3$ \& 4. We find the vast majority of measurements have a one-to-one agreement (\textit{black dashed line}) with a 95\% confidence interval of $(-0.28, +0.42)$ dex (\textit{shaded black region}). We find negligible systematic offset between the two assessments.}
\label{fig:comp_mass}
\end{figure}

\begin{figure*}
	\centering
	\includegraphics[width=\textwidth]{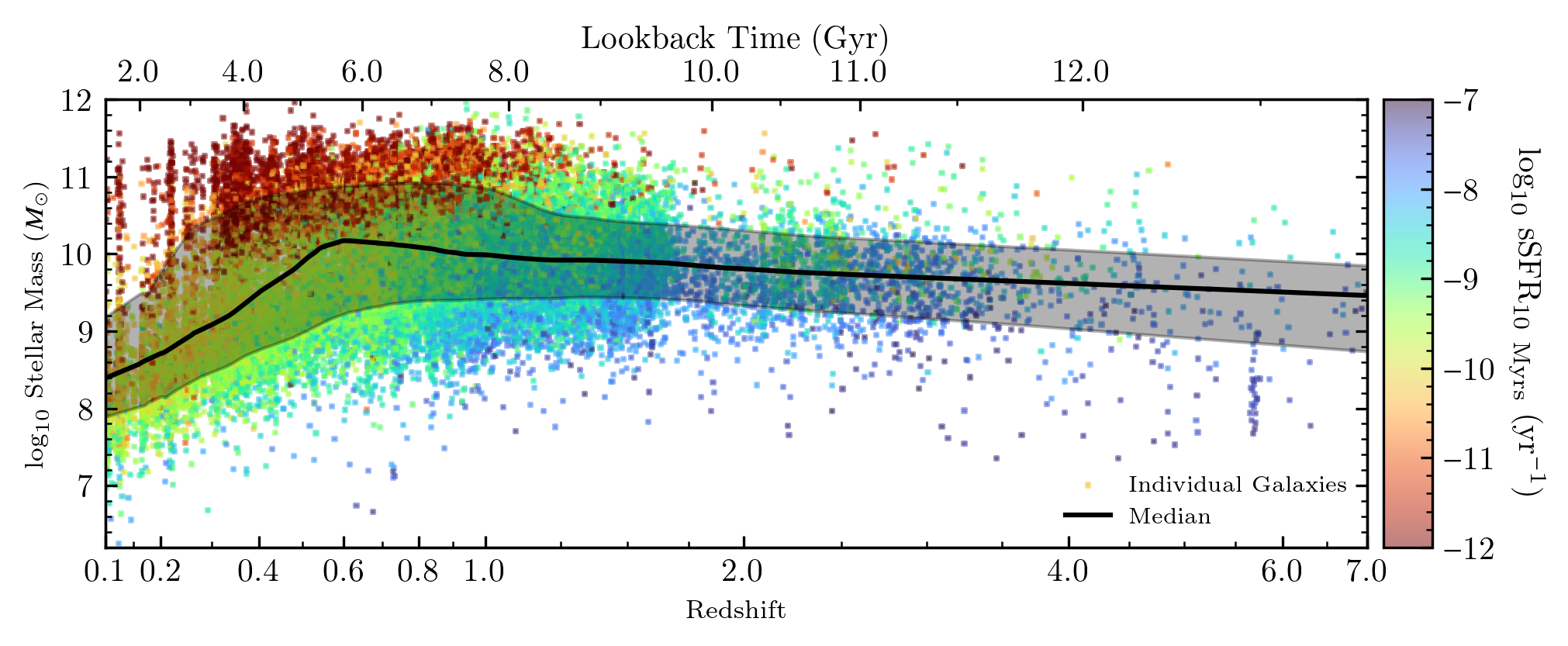}
	\caption{The stellar mass (\cigale) -- redshift distribution of high \qf~sources within the compilation color coded by their \cigale-determined specific SFR on a 10 Myr timescale. The median and $1\sigma$ scatter in stellar mass are shown as the \textit{black solid line} and \textit{black shaded region}, respectively. Quiescent (low sSFR) systems are primarily found at $z < 1.25$ with $>10^{10}$\,\msol, while bursty star-forming (high sSFR) systems are primarily at $z > 2$ extending to the highest redshifts in the compilation. The median stellar mass of the compilation is $< 10^{9.4}$ \msol~at $z < 0.4$ with the distribution being bimodal with a high-mass and a low-mass peak. The median mass peaks at $z\sim0.6$ and steadily decreases to $z\sim7$ from $\sim10^{10}$\,\msol~to $10^{9.4}$\,\msol, respectively.}
	\label{fig:mass_z}
\end{figure*}

Figure \ref{fig:completeness_spatial} shows the spatial completeness without limiting \qf~is mostly $> 50\%$ throughout with $>80$\% within the central area of COSMOS. This is mainly due to the many spectroscopic follow-up programs of galaxies identified within \textit{HST}/CANDELS \citep{Grogin2011,Koekemoer2011} region as shown by the \textit{pink} (ACS) and \textit{red} (WFC3/IR) footprints overlaid on Figure \ref{fig:completeness_spatial} as well as the DESI EDR \citep{Adame2024}, DR1 \citep{AbdulKarim2025}, and DR2-COSMOS \citep{Ratajczak2025} datasets. Limiting to high \qf, we find that the CANDELS/COSMOS region has $>60\%$ completeness, while moving further away results in decreasing completeness ($\sim 30 - 40\%$). This is expected given that the CANDELS program focused on deep, spatially-resolved imaging of galaxies serving as prime objects for spectroscopic follow-up. We do note other smaller regions that have elevated spectroscopic completeness such as one just southwest (RA $ \sim 150$ deg, Dec.\ $ \sim 2$ deg) of CANDELS that has $\gtrsim 50\%$ spectroscopic completeness ($\textrm{\qf} = 3 - 4$).  

Overall, Figure \ref{fig:completeness_spatial} highlights how much spectroscopic focus has been placed on the central regions of COSMOS given the wealth of space-based, spatially-resolved imaging to accompany spectroscopic studies. At the same time, wide-field spectroscopic surveys such as DESI have helped to increase the spatial completeness to $\sim 30 - 45\%$ further from the central regions. However, Figure \ref{fig:completeness_spatial} highlights the need to still spectroscopically explore further away 
\nolinebreak~from the central region as well as push towards fainter magnitudes where the spatial completeness decreases throughout the field. This is already changing with the advent of \textit{JWST} programs such as COSMOS-Web (PIs: J. Kartaltepe \& C. Casey, \citealt{Casey2023}; \textit{orange} footprint in Figure \ref{fig:completeness_spatial}), COSMOS-3D (PI: K. Kakiichi), as well as other programs such as CLUTCH, an HST Multi-Cycle Treasury program to obtain UV-NIR imaging over the COSMOS-Web area (PI: J. Kartaltepe), that will motivate future ground- and space-based spectroscopic observations populating regions currently lacking in spectroscopic follow-up. Programs such as COSMOS-3D, CAPERS (PI: M. Dickinson), PASSAGE (PI: M. Malkan), POPPIES (PI: J. Kartaltepe), and SAPPHIRES (PI. E. Egami) are already underway to fill in some of these gaps. Large wide-field, ground-based spectroscopic campaigns such as Subaru/PFS \citep{Greene2022}, WAVES/4MOST \citep{Driver2019}, and VLT/MOONS/MOONRISE \citep{Maiolino2020} will also significantly increase the spatial completeness throughout COSMOS.

\begin{figure*}
	\centering
	\includegraphics[width=\textwidth]{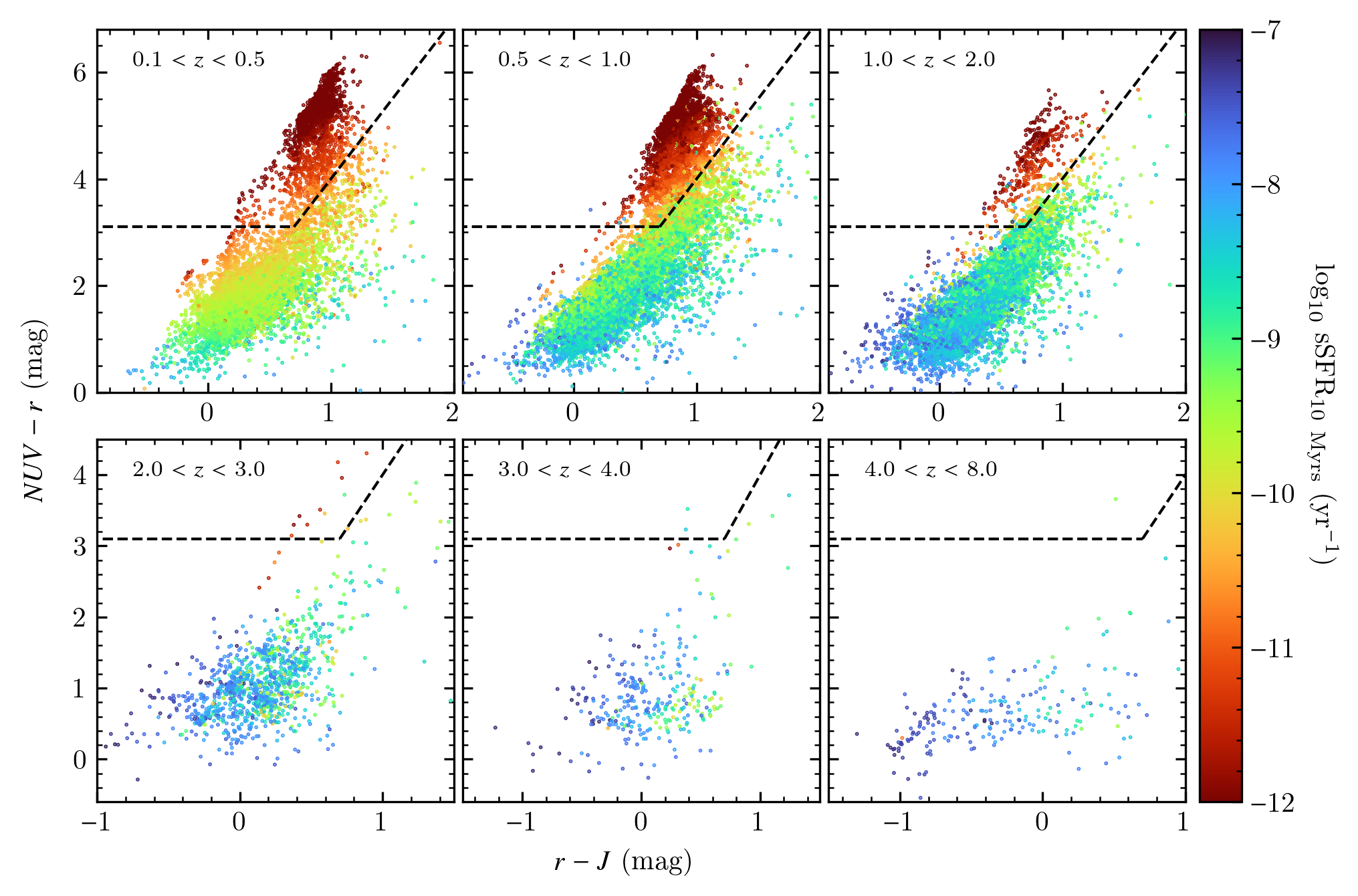}
	\caption{Rest-frame $NUV - r$ vs $r-J$ colors color-coded by sSFR within a 10 Myr timescale. The \textit{black dashed line} represents the quiescent (above) and star-forming (below) galaxy separation as defined by \citet{Ilbert2013}. The compilation contains numerous $z < 2$ passive/quiescent galaxies with sSFR$ < 10^{-11}$ yr$^{-1}$ as well as star-forming galaxies with many within the blue $NUVrJ$ loci corresponding to young star-forming galaxies with sSFR $> 10^{-8}$ yr$^{-1}$. At $z > 2$, we find the vast majority of galaxies reside within blue $NUVrJ$ colors and have sSFR reaching $10^{-7}$ yr$^{-1}$. Only a few galaxies fall within the passive galaxy selection area at $z > 2$; however, we do note several `star-forming' galaxies exhibiting redder $NUV - r$ colors and low sSFR which could be galaxies transitioning into a quiescent state (e.g., post-starburst galaxies).}
	\label{fig:NUVrJ}
\end{figure*}

\subsection{Stellar Mass Assessments}

As described in \S\ref{sec:SED}, we use both \cigale~and \lephare~to reassess stellar mass measurements of galaxies with \qf$= 3$ \& 4 and redshifts fixed to the best spectroscopic redshift in the compilation. Throughout this work, we refer to the stellar mass and associated physical properties as determined from \cigale~but also include \lephare~as a secondary SED fitting measurement, as well as for consistency, as past COSMOS catalogs have used \lephare~photo-$z$ for SED measurements. Figure \ref{fig:comp_mass} shows the comparison between stellar mass determined using both approaches, where we find $95\%$ of measurements are in agreement within $-0.3$ to $+0.4$ dex.

The range of stellar masses in the compilation along with the median stellar mass and $1\sigma$ scatter at a given redshift are shown in Figure \ref{fig:mass_z}, where we find a stark difference between the $z< 1$ and $z > 1$ populations. The typical stellar mass at $z < 1$ ranges from 10$^{8.4}$ to $10^{10}$ \msol~with a large scatter ranging from 10$^8$ to $10^{11}$ \msol~highlighting the diverse population at low-$z$ within the compilation. Figure \ref{fig:mass_z} includes each individual galaxy color-coded with its sSFR where we find that the $z < 1$ subset incorporates both massive galaxies with low sSFR ($<10^{11}$ yr$^{-1}$) and low-mass, high sSFR galaxies ($> 10^{-9}$ yr$^{-1}$). Several dwarf-like systems (e.g., $< 10^8$ \msol) are also included in the compilation that have relatively high sSFR reaching as high as $10^{-7}$ yr$^{-1}$, which represents a subset of bursty low-$z$ star-forming galaxies.

Selection effects become increasingly noticeable at $z > 1$ as the median stellar mass decreases from $\sim10^{10}$ \msol~($z\sim1$) to 10$^{9.4}$ \msol~($z\sim7$) and the scatter tightens from $z\sim1$ to $1.5$ and is relatively constant with increasing redshift. Galaxy populations in the compilation at $z > 1$ are heavily dominated by low-mass, high sSFR star-forming galaxies. For example, a noticeable cluster of $<10^9$ \msol~galaxies at $z \sim 5.7$ with sSFR $\sim 10^{-7}$ to 10$^{-8}$ yr$^{-1}$ is seen in Figure \ref{fig:mass_z}, which corresponds to spectroscopic confirmations of narrow band-selected \lya~emitters mainly within 10K-DEIMOS \citep{Hasinger2018} and M2FS \citep{Ning2020}. We note the former is based on a compilation of several COSMOS DEIMOS programs, including a follow-up of narrowband-selected \lya~emitters (e.g., \citealt{Taniguchi2015}) and Lyman Break Galaxies (LBGs) selected via dropout technique. Indeed, at $z > 3.5$, the majority of spectroscopic redshift measurements in the compilation are based on \lya~emission observed within observer-frame optical wavelengths and bias the population towards low-mass, dust-free young galaxies. Future additions that will include the latest \textit{JWST}/NIRspec measurements could help in diversifying the high-$z$ galaxy population in the compilation where studies are finding massive, quiescent galaxies at $z  > 3$ (e.g., \citealt{Nanayakkara2024,Sato2024,Slob2024}).

\begin{figure*}
	\centering
	\includegraphics[width=\textwidth]{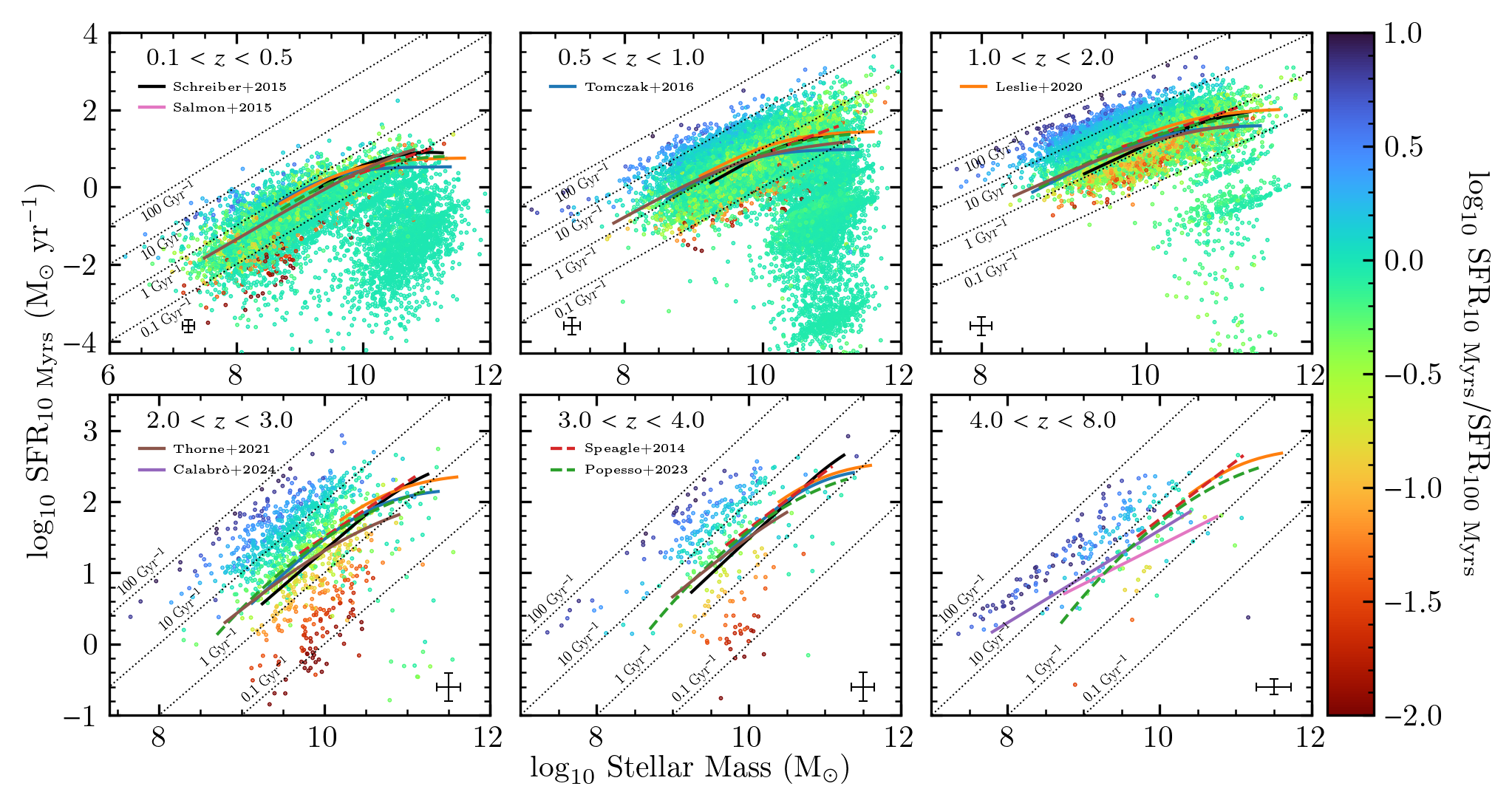}
	\caption{The SFR -- stellar mass correlation for different redshift subsets color-coded by their 10-to-100 Myr SFR ratios. Included are observed \citep{Schreiber2015,Salmon2015,Tomczak2016,Leslie2020,Thorne2021,Calabro2024} and compilation \citep{Speagle2014,Popesso2023} measurements limited to the stellar mass ranges of the sample (model) limits. The compilation shows the massive, quiescent populations at $>10^{10}$ \msol~up to $z \sim 2$ consistent with a turnoff in the SFR -- stellar mass correlation. sSFR $> 10$ Gyr$^{-1}$ systems also corresponding to $\log_{10} \textrm{SFR}_\textrm{10Myr}/\textrm{SFR}_\textrm{100Myr} > 0$ correspond to starburst outliers in the SFR -- stellar mass correlation with the vast majority of the $z > 4$ sources in the compilation falling within this regime. Overall, the compilation at $z < 4$ is consistent with past SFR -- stellar mass correlation measurements with the spread attributed to variations in star-formation activity and histories.}
	\label{fig:SFR_Mass}
\end{figure*}

\subsection{Quiescent, Star-Forming, \& Bursty Galaxy Populations}

In this section, we take a closer look at the various galaxy populations that encompass the compilation using two classical approaches: $NUVrJ$ rest-frame color-color diagnostics \citep{Ilbert2010,Ilbert2013} and the SFR -- stellar mass correlation (SFR `main sequence'; e.g., \citealt{Noeske2007,Daddi2007}). We first focus on the former where we use the rest-frame \textit{GALEX}/$NUV$, Subaru HSC/$r$, and Paranal VISTA/$J$ photometry that was measured by convolving the respective filters with the best-fit SED during the \cigale~fitting process. We follow \cite{Ilbert2013} in using $NUV - r$ and $r - J$ colors to classify galaxies as quiescent or star-forming, where $NUV - r > 3.1$ and $NUV - r > 3 (r - J) +1$ are considered quiescent. This is similar to the commonly used $UVJ$ diagnostic \citep{Williams2009}; however, $NUV - r$ is found to better trace recent star formation activity compared to $U - V$ where $NUV$ is sensitive to stellar populations with a light-weighted age of $10^8$ yr and $r$ traces $> 10^9$ yr stellar populations \citep{Arnouts2007,Martin2007}.

Figure \ref{fig:NUVrJ} shows the rest-frame $NUVrJ$ colors at different redshifts, with each galaxy color-coded by its sSFR. We find clear, distinct populations of galaxies at varying redshifts making up the compilation. Numerous quiescent galaxies exhibiting low sSFR reaching $10^{-12}$ yr$^{-1}$ are present from $z \sim 0 - 3$ with the majority of them at $z < 1$. These are primarily high-mass galaxies as shown in Figure \ref{fig:mass_z}. However, their numbers significantly decrease progressing from $z \sim 1$ to higher redshifts where by $4 < z < 8$ we find only a single object with a spectroscopic redshift that could be a quiescent galaxy, which is mainly due to a combination of selection bias within the compilation and the decrease in quiescent fractions with increasing redshift. 


Populations of star-forming galaxies are also found to be present at all redshifts with several key differences. The $z > 3$ population is mostly clustered around a loci in $NUV - r \sim 0 - 2$ mag and $r - J \sim 0$ mag that signifies a population consisting of young stellar populations (recent star formation) and relatively little dust (blue $r - J$).  The loci also corresponds to galaxies that have sSFR $> 10^{-8}$ yr$^{-1}$ signifying $< 100$ Myr mass-doubling times. This is most likely a byproduct of selection functions in the high-$z$ part of the compilation being heavily dependent on the confirmation of \lya~emitters within observer-frame optical wavelengths. At $z < 3$, we start to note both a population of star-forming galaxies at the same loci as well as an extension to redder $r - J$ colors, signifying the inclusion of dusty star-forming galaxies. This is due to galaxies observed with strong rest-frame optical emission lines (e.g., \halpha, \oiii, \oii) within observer-frame optical and near-IR spectroscopy which are less susceptible to dust extinction compared to rest-frame UV lines (e.g., \lya). In effect, dusty star-forming galaxies would be included in the compilation at $z < 3$, given how the selection was done for follow-up observations. Post-starburst galaxies are also within the compilation mostly at $z < 2$ showing similar $0 < r - J < 0.5$ mag and elevated $NUV - r$ suggesting a population of galaxies potentially transitioning from the young, dust-free star-forming galaxy loci towards the mature, passive galaxy classification.

Figure \ref{fig:SFR_Mass} shows the SFR -- stellar mass correlation (`main sequence') for the same redshift ranges in Figure \ref{fig:NUVrJ} and highlights the diverse range of galaxies in the compilation at low-$z$. Overlaid on each panel are measurements from previous studies \citep{Speagle2014,Schreiber2015,Salmon2015,Tomczak2016,Thorne2021,Popesso2023,Calabro2024} limited to their respective stellar mass and redshift ranges. Overall, the majority of sources in the compilation at $z < 4$ fall along past studies of the SFR -- stellar mass correlation with the scatter highlighting the diverse range of galaxies in the compilation. At $z < 1$, we find many $>10^{10}$ \msol~ galaxies falling well below the main sequence implying a quiescent, passive system. These sources also make up the low sSFR quiescent cloud seen in Figure \ref{fig:NUVrJ}. Several studies also capture this turnover in the SFR -- stellar mass correlation (e.g., \citealt{Whitaker2014,Lee2015,Schreiber2015,Tomczak2016,Leslie2020,Cooke23, Popesso2023}). 

At $4 < z < 8$, we find many galaxies lie well above the SFR -- stellar mass correlation signifying populations of starburst galaxies. This is also evident with the color-coding in Figure \ref{fig:SFR_Mass}, which shows the ratio between star formation rates as measured on a 10 Myr and 100 Myr timescale via \cigale. All sources above the correlation at $z > 4$ have $\log_{10} \textrm{SFR}_\textrm{10Myr}/\textrm{SFR}_\textrm{100Myr} > 0$ implying recent burst of star-formation activity and is supported by the elevated sSFR suggesting mass-doubling time scales of $< 100$ Myr. As described above, the high-$z$ regime of the compilation is biased towards \lya~emitters which are known for their young stellar populations (recent star-formation activity; e.g., \citealt{Malhotra2002,Ono2010,Santos2020}) which explains the many sources lying well above the SFR -- stellar mass correlation.

The compilation also includes numerous $1 < z < 4$ starburst galaxies exhibiting sSFR $> 10^{-8}$ yr$^{-1}$ and $\textrm{SFR}_\textrm{10Myr}/\textrm{SFR}_\textrm{100Myr}$ reaching up to 10. However, the number of starburst galaxies starts to decrease at $z < 2$ with no galaxy in the compilation having a sSFR $> 10^{-8}$ yr$^{-1}$ at $z < 0.5$ (although this is partly due to selection effects; see \S\ref{sec:SOMs}). This also falls in line with the gradual decrease in the cosmic star formation history (e.g., \citealt{Madau2014,Khostovan2015,Zavala2021}) and cosmic specific star formation rate (e.g., \citealt{Speagle2014,Faisst2016,Khostovan2024}). Overall, the compilation includes a wide range of galaxies ranging from bursty to typical star-forming galaxies and massive, quiescent galaxies.

\subsection{Diversity of Environments}

\begin{figure}
    \centering
        \includegraphics[width=\columnwidth]{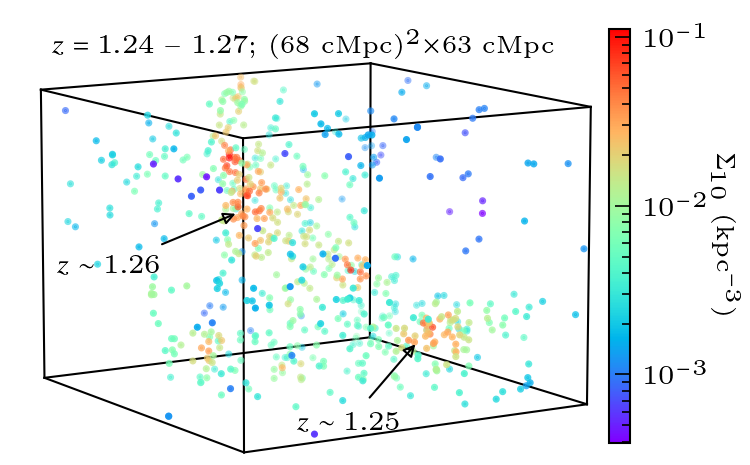}
        \includegraphics[width=\columnwidth]{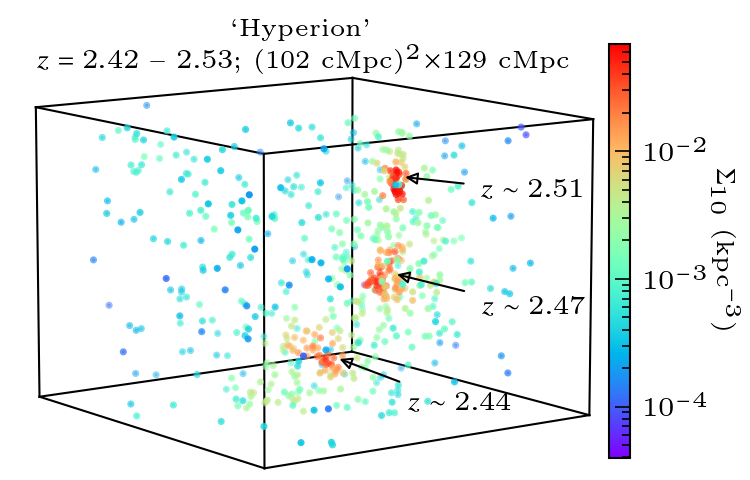}
    \caption{Examples of large-scale structure features in the compilation traced by the 10$^\mathrm{th}$ nearest neighbor algorithm. Each example encompasses the full COSMOS angular area with only a cut in redshift space. \textit{Top panel} shows a thin redshift slice at $z \sim 1.24$ -- $1.27$ (63 cMpc LOS distance), where we find two overdense regions separated by $\sim 40$ cMpc. \textit{Lower panel} shows the extensively studied `Hyperion' system \citep{Cucciati2018}, which we find to be composed of 3 overdense regions. Such 3D visualizations of LSS are only possible given the many spectroscopic redshift measurements within the compilation and demonstrate how the compilation can be used to effectively visualize and study large-scale structure and its effects on galaxy evolution. An animated version of this figure is available online and provides a full 360 degree rotation around the center of the box allowing for a full visualization of large-scale structure.}
    \label{fig:LSS_examples}
\end{figure}

The large number of sources spread out with a wide, contiguous area and high-quality spectroscopic redshifts in the compilation allows us to visualize 3D large-scale structure in the field, highlighting the diverse range of environments that galaxies reside in, such as field, filaments, and clusters. This also enables the compilation to be used in studies focused on how environment (position of a galaxy within large-scale structure) can affect the underlying evolution of galaxy populations.

There are many different density measurement approaches to map out large-scale structures. These include Voronoi Tesselation (e.g., \citealt{Scoville2013, Darvish2015,Tomczak2016,Lemaux2017,Brinch2023,Hung2024}), Kernel Density Estimation (e.g., \citealt{Darvish2014,Darvish2015,Chartab2020,Brinch2023,Taamoli2024}), Optimal Filtering (e.g., \citealt{Bellagamba2011,Maturi2019,Toni2025}), Friends-of-Friends Group Finding (e.g., \citealt{Knobel2012}) , and Nearest Neighbor algorithm (e.g., \citealt{Cooper2008,Sobral2011,Sillassen2022,Sillassen2024}). For the purposes of our simple visual demonstration, we measure the densities using a 10$^{\textrm{th}}$ Nearest Neighbor approach. This analysis is limited only to those that have \qf$ = 2 - 4$ and $12 - 14$ with the inclusion of the \qf$ = 2,12$ flags to increase the sample size to better trace large-scale structure while also not significantly increasing the uncertainties due to poorly constrained spectroscopic redshifts. We convert from angular -- $z$ space into 3D comoving space. This is done by first centering the angular space on the center of COSMOS (RA$ = 150$ deg, Dec. $ = 2.2$ deg), then calculating the traverse comoving separation from the center. The line-of-sight comoving distance is measured using the spectroscopic redshift forming the $z$ axis. We then calculate the local density for each galaxy as:
\begin{eqnarray}
	\Sigma_{10} = \frac{11}{\frac{4}{3} \pi r_{10}^3},
\end{eqnarray}
where $\Sigma_{10}$ is the local density around the reference galaxy and the nearest 10 neighbors within a spherical region of size $r_{10}$ corresponding to the distance of the 10$^{\textrm{th}}$ neighbor (maximum distance). Note this does not include any corrections for completeness and edge effects and is only meant to be a visual demonstration of the variety of environments that are traced within the compilation. Careful considerations need to be made for proper density measurements.

Figure \ref{fig:LSS_examples} shows two interesting examples of large-scale structure within the compilation. Both boxes are limited in width to 1 deg in both RA and Dec., which corresponds to 68 cMpc and 102 cMpc for the \textit{top} and \textit{bottom} panels, respectively. The $z$ axis of each panel corresponds to increasing redshift from bottom to top. For each box, we also label the redshift range shown as well as the comoving dimensions where $x$ and $y$ are of same size.

The \textit{bottom} panel of Figure \ref{fig:LSS_examples} corresponds to the large and extended overdense region known as `Hyperion,' first fully identified as being a super-protocluster with multiple components by \citet{Cucciati2018} within VUDS \citep{LeFevre2015,Tasca2017} and $z$COSMOS-Bright \citep{Lilly2007} and Deep (PI: Simon Lilly; \citealt{Diener2015}) spectra. Previous studies also traced this structure via \lya~emitters (e.g., \citealt{Chiang2015, Huang2022}), \lya-forest tomography (e.g., \citealt{Lee2016,Newman2020}), CO-emitting galaxies (e.g., \citealt{Wang2016,Champagne2021}), and submillimeter star-forming galaxies (e.g., \citealt{Casey2015}). Figure \ref{fig:LSS_examples} also shows the larger density peaks correlated with 3 distinct clumps at $z \sim 2.44$, $2.47$, and $2.51$, corresponding to a width of $\sim 80$ cMpc. Elevated $\Sigma_{10}$ in between each overdense peak could potentially be connecting filamentary structure.

The \textit{top} panel of Figure \ref{fig:LSS_examples} shows two overdense regions at $z \sim 1.25$ (RA = 149.8 deg, Dec.\ = 2.05 deg) and $z \sim 1.26$ (RA = 150.1 deg, Dec.\ = 2.39 deg) separated by $\sim 40$ cMpc not previously identified in the literature. Based on our simple assessment of $\Sigma_{10}$, we find that the two overdensities seem to be surrounded by somewhat elevated $\Sigma_{10}$ that could also signify both overdensities are peaks in a much larger structure. A closer look at the $z \sim 1.26$ overdensity shows an extension of $\Sigma_{10} \sim 0.01$ kpc$^{-3}$, extending down halfway to the $z \sim 1.25$ peak providing some suggestion that the two overdensities are connected. We note that both overdensities were also identified via a separate large-scale structure tracing approach that used our compilation \citep{Toni2025}.

These are just two examples of overdensities that are part of the compilation and we note that several other overdensities are present, especially towards the low-$z$ end of the compilation. Note that in both examples, we not only have peaks corresponding to regions where galaxies are spatially clustered with one another but also regions with $\Sigma_{10} < 10^{-3}$ kpc$^{-3}$ that are sparsely populated, highlighting how the compilation has a wide, diverse range of environments and can be used for detailed environmental (large scale structure) studies (e.g., comparing physical properties of galaxies residing in overdense regions and within sparsely populated fields). 

\section{Application Examples}
\label{sec:applications}

The compilation has many practical applications for both calibration and scientific analyses. As listing and describing each would go beyond the scope of this paper, we only focus on two main applications: validation of photometric redshifts in surveys (\S\ref{sec:photoz_validation}) and using self-organizing maps to infer which galaxy populations currently lack spectroscopic follow-up thereby enabling new science and future observing campaigns (\S\ref{sec:SOMs}).

\begin{table*}
	\centering
	\begin{tabular*}{\textwidth}{@{\extracolsep{\fill}} l l c c c c}
		\hline
		Catalog & Reference & $N_{obj}$ & $\sigma_\textrm{NMAD}$ & $\eta$ & $b$ \\
		&           &           &      &  (\%)  &      \\
		\hline
		\multicolumn{6}{l}{\textbf{Galaxies}}\\
        COSMOS2009 & \cite{Ilbert2009} & 30329 & 0.012 & 5.0 & -0.001\\
        COSMOS2015 & \cite{Laigle2016} & 37083 & 0.013 & 8.6 & -0.001\\
        COSMOS2020 (Classic) & \cite{Weaver2022} & 25476 & 0.011 & 3.2 & -0.003\\
        COSMOS2020 (Farmer) & \cite{Weaver2022} & 23631 & 0.012 & 3.4 & 0.000\\
		\\
		\multicolumn{6}{l}{\textbf{X-ray AGN}}\\
        COSMOS2015 & \cite{Laigle2016} & 869 & 0.027 & 20.5 & -0.003\\
        COSMOS2020 (Classic) & \cite{Weaver2022} & 1704 & 0.024 & 8.5 & -0.009\\
        COSMOS2020 (Farmer) & \cite{Weaver2022} & 1207 & 0.030 & 10.2 & -0.009\\
        \textit{Chandra}/COSMOS-Legacy (All) & \cite{Marchesi2016} & 595 & 0.014 & 10.3 & -0.001\\
        \textit{Chandra}/COSMOS-Legacy (Type 1) & \cite{Marchesi2016} & 429 & 0.012 & 10.7 & -0.000\\
        \textit{Chandra}/COSMOS-Legacy (Type 2) & \cite{Marchesi2016} & 166 & 0.020 & 9.0 & -0.005\\
		\hline
	\end{tabular*}
	\caption{Photometric Redshift quality assessment of past COSMOS catalogs using our spec-$z$ compilation. All photometric redshifts are measured using \lephare~with the exception of \cite{Marchesi2016} which used \lephare~with an AGN template library described in \cite{Salvato2011}. Spectroscopic matches used to validate photometric redshifts are restricted to $Q_f = $ 3 -- 4 and 13 -- 14. For each catalog shown, we include the number of galaxies in the sample ($N_{gal}$), the normalized median absolute deviation ($\sigma_\textrm{NMAD}$; Equation \ref{eqn:NMAD}), the outlier fraction ($\eta$; Equation \ref{eqn:eta}), and bias ($b$; Equation \ref{eqn:bias}).}
	\label{table:photoz}
\end{table*}

\begin{figure*}
	\centering
	\includegraphics[width=\textwidth]{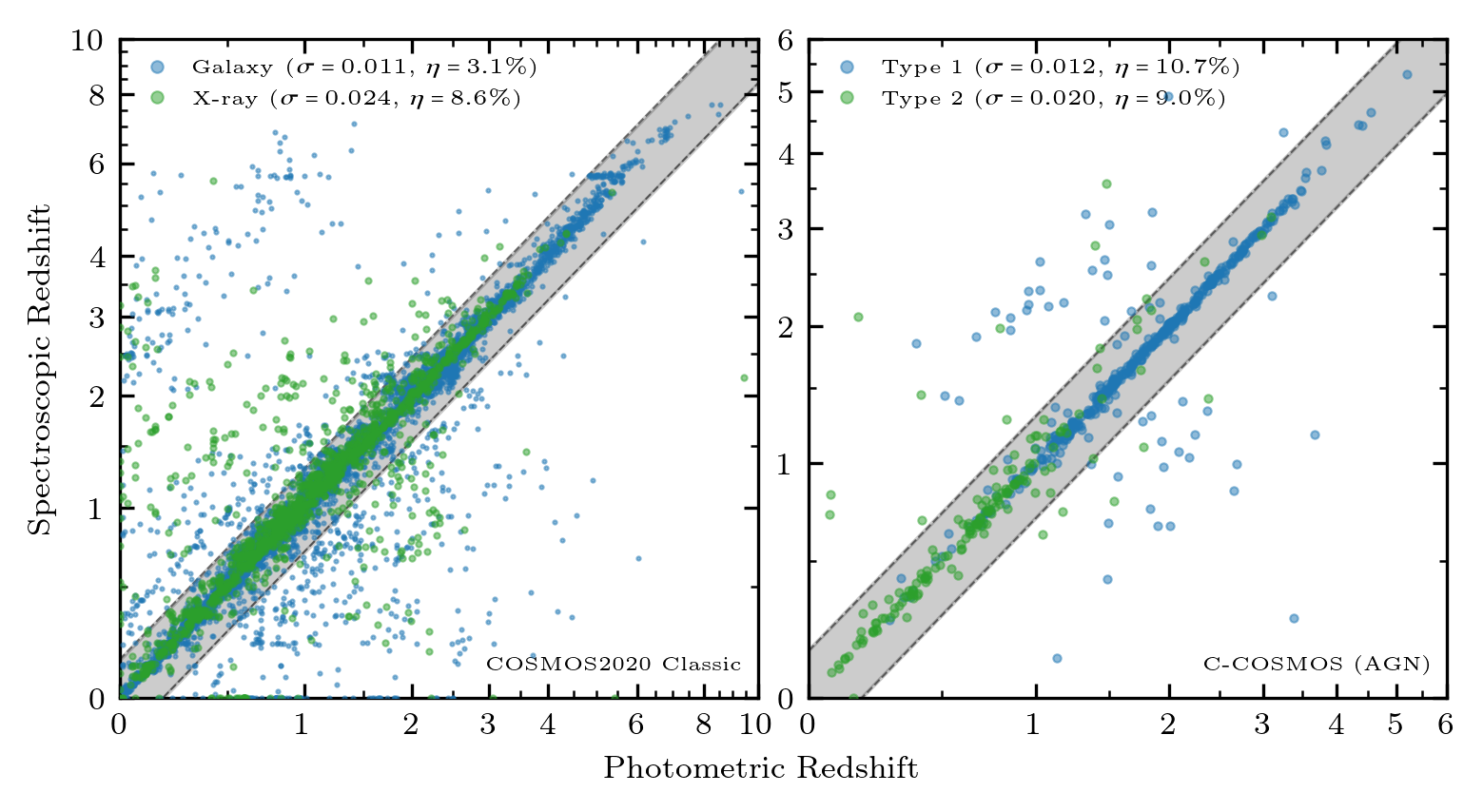}
	\caption{Comparison between photometric and spectroscopic redshifts for COSMOS2020 Classic (\textit{left}) and \textit{Chandra}/COSMOS-Legacy (\textit{right}). The \textit{grey} region corresponds to the condition $\lvert (z_\textrm{phot} - z_\textrm{spec})/(1+z_\textrm{spec}) \rvert < 0.15$ with sources outside this region making up the outlier fraction ($\eta$). The corresponding NMAD ($\sigma$) is shown for each subsample. Both COSMOS2020 and C-COSMOS use {\sc LePHARE} for photometric redshift measurements with the difference that C-COSMOS specifically uses AGN templates \citep{Salvato2011,Marchesi2016} and COSMOS09 photometry \citep{Ilbert2009} while COSMOS2020 photo-$z$ are measured using both galaxy \& AGN templates (whichever has the best $\chi^2$). Both data sets shows the need for improved AGN templates to reduce outlier fractions which the compilation could serve as a valuable calibration data set in such an endeavour.}
\label{fig:validation}
\end{figure*}

\subsection{Validation of Photometric Redshifts}
\label{sec:photoz_validation}

Any galaxy imaging survey requires proper assessment of each galaxy's redshift in order for the data set to be used for a multitude of science cases. Typically this is done by using all available multi-wavelength photometry for each galaxy \citep[see][for a review]{Salvato2019} and using at least one of the many template-based photo-$z$ codes in the literature such as \lephare~\citep{Arnouts1999,Ilbert2006} and \texttt{EaZY} \citep{Brammer2008}. However, photo-$z$ measurements are limited by the available photometry (e.g., not enough to constrain SEDs in template fitting) and limited grid space in physical parameters that generate the templates used for photo-$z$ determination. Therefore, it is crucial to validate photo-$z$ measurements by testing how well the measured photo-$z$ matches with known spectroscopic redshifts on a population level. Compilations such as the one presented in this paper are valuable resources in testing the reliability and validity of measured photometric redshifts (e.g., \citealt{Laigle2016,Battisti2019,Weaver2022}). In this section, we will demonstrate how the compilation can be used to validate photo-$z$ measurements from various versions of the COSMOS catalogs, highlight any systematics, and also gauge the reliability of redshift measurements for different galaxy population subsets (e.g., bright vs faint galaxies).

We first define the diagnostics that we will use to assess the photo-$z$ validation. The systematic offset between the photometric and spectroscopic redshift is measured as the redshift bias, $b$, defined as:
\begin{eqnarray}
	b =& \textrm{median}\Big(\frac{z_\textrm{phot} - z_\textrm{spec}}{1+z_\textrm{spec}}\Big)\
	\label{eqn:bias}
\end{eqnarray}
with $z_\mathrm{phot}$ and $z_\mathrm{spec}$ being the photometric and spectroscopic redshfits, respectively. A non-zero $b$ signifies a systemic offset in $z_\mathrm{phot}$ relative to $z_\mathrm{spec}$ for the sample as a whole. The precision of $z_\mathrm{phot}$ is assessed by using the normalized median absolute deviation (NMAD; \citealt{Hoaglin1983}) defined as:
\begin{eqnarray}
	\sigma_\textrm{NMAD} = 1.4821 \times \textrm{median}\Bigg(\frac{\lvert \Delta z - \textrm{med}(\Delta z)\rvert}{1+z_\textrm{spec}}\Bigg),
	\label{eqn:NMAD}
\end{eqnarray}
where $\Delta z = z_\textrm{phot} - z_\textrm{spec}$ which has been found to be unaffected by outliers (e.g., \citealt{Ilbert2006}). The outlier fraction, $\eta$, represents the fraction of galaxies for which $z_\mathrm{phot}$ deviates from $z_\mathrm{spec}$ by:
\begin{eqnarray}
	\Big\lvert\frac{z_\textrm{phot} - z_\textrm{spec}}{1+z_\textrm{spec}}\Big\rvert > 0.15,
	\label{eqn:eta}
\end{eqnarray}
where the criteria of $0.15$ is based on past photo-$z$ validation tests (e.g., \citealt{Hildebrandt2012,Weaver2022}).

For all validation tests, we cross-match each respective photometric catalog with the compilation limited to only those that have high quality spectroscopic redshifts (\qf$=3, 4, 13, 14$). All photometric redshifts tested are based on \texttt{LePHARE} and we name each column from the original released catalogs that were used in the photo-$z$ validity tests. We refer the reader to the references listed in Table \ref{table:photoz} for specific details on the parameters adopted for \lephare~in each case. For both COSMOS2020/Classic and Farmer catalogs, we use the best-fit photometric redshift (\texttt{lp\_zbest}) in the case where the source is classified as a galaxy based on its best-fit template (\texttt{lp\_type} $ = 0$). If the best-fit template is associated with an X-ray source (\texttt{lp\_type} $ = 2$), then we use the AGN-template (see Table 3 of \citealt{Salvato2009} and \citealt{Salvato2011}) determined photometric redshift (\texttt{lp\_zq}) only under the condition that the AGN-template $\chi^2$ is less than the galaxy-template $\chi^2$ (\texttt{lp\_chiq} $ < $ \texttt{lp\_chi2\_best}). In the case of COSMOS2015, we use the best-fit $z_{phot}$ (\texttt{ZMINCHI2}) under the condition that \texttt{TYPE} $ = 0$ (galaxy-template) and only use the AGN-template $z_{phot}$ (\texttt{ZQ}) under the same condition as in COSMOS2020 (\texttt{CHIQ} $<$ \texttt{CHI2\_BEST}). For COSMOS2009, we only test the photo-$z$ (\texttt{zp\_best}) from the best-fit galaxy templates (\texttt{type} $ = 0$). Lastly, we also perform validity tests of the photo-$z$ measurements for \textit{Chandra}/COSMOS \citep{Marchesi2016} sources limited to only those corresponding to Type 1 (\texttt{phot\_type} $ = 1$) and Type 2 (\texttt{phot\_type} $ = 2$) AGN.

Table \ref{table:photoz} shows the photo-$z$ validation tests for all the catalogs mentioned above separated based on the redshifts being measured from galaxy and X-ray AGN templates. Overall the bias in each survey is consistent with zero with only the COSMOS2020 catalogs for the X-ray AGN template measured $z_\mathrm{phot}$ having a $b \sim -0.01$, which is still quite negligible. This suggests that past photo-$z$ assessments do not suffer from any systematic offsets on a population level. Figure \ref{fig:validation} also shows how the COSMOS2020 Classic and \textit{Chandra}/COSMOS catalogs have no clear systematic offset from a one-to-one relation between the photometric and spectroscopic redshifts. For all surveys, $\sigma_\textrm{NMAD}$ is consistently around $\sim 0.01$ for the galaxy-template $z_\mathrm{phot}$ suggesting that the inclusion of newer and deeper data since the COSMOS2009 catalog has not changed the precision of photo-$z$ measurements. However, the outlier fraction has improved significantly, with COSMOS2015 having the highest at $8.6\%$ while the COSMOS2020 catalog has $3.2$ and $3.4\%$ for Classic and Farmer, respectively. This is also highlighted in Figure \ref{fig:validation}, which shows the vast majority of photo-$z$ measurements are within the 0.15 criteria (Equation \ref{eqn:eta}; \textit{black} shaded region in Figure \ref{fig:validation}). The outliers are found to be primarily $z < 1$ galaxies that are spectroscopically confirmed to be $z > 2$ galaxies corresponding to the Lyman/Balmer break degeneracy. Inclusion of near-IR photometry is very useful in breaking this degeneracy (e.g., \citealt{Battisti2019}).

The validity of photo-$z$ measurements using AGN-based templates is found to be less reliable compared to the galaxy-template photo-$z$ measurements. Table \ref{table:photoz} shows the validity statistics for the AGN-template $z_\mathrm{phot}$ measurements, where we find no significant bias; however, $\sigma_\textrm{NMAD}$ is $\sim 2 - 3 \times$ higher than what is found for galaxy-templates except for the full \textit{Chandra}/COSMOS-Legacy sample and its Type 1 AGN subset ($\sigma_\textrm{NMAD} \sim 0.01$). The outlier fractions are also significantly higher, with the COSMOS2015 catalog having the highest at $\sim 20\%$. The COSMOS2020 catalogs have similar $\eta \sim 10\%$ as the \textit{Chandra}/COSMOS measurements; although, this is $\sim 3\times$ higher than $\eta$ from galaxy-based templates. Figure \ref{fig:validation} shows how the photo-$z$ compares with spec-$z$ for Type 1 and Type 2 AGN, where we find no systematic offsets from a one-to-one relation but find elevated $\eta$. This could suggest that improved AGN templates are needed to better reduce outlier fractions. In this regard, the compilation could potentially be used to derive such improved templates.

Figure \ref{fig:photoz_validation_imag} shows the photo-$z$ validation statistics for the COSMOS2020 Classic, Farmer, and \textit{Chandra}/COSMOS catalogs as a function of HSC $i$ magnitude. As we also found for the full population photo-$z$ validity tests, the galaxy template-based photo-$z$ measurements show the best results in terms of lower $\sigma_\textrm{nmad}$ and $\eta$ at each given $i$ magnitude. In all three catalogs, we find that the precision decreases and outlier fractions increase significantly towards fainter magnitudes. The outlier fractions of COSMOS2020 Classic (galaxy template-based) reach $>10\%$ by $i \sim 25$ mag and $\sim 30\%$ by $i \sim 27$ mag. This is close to the 3$\sigma$ limit (27.6 mag; Table 1 of \citealt{Weaver2022}) such that it is expected $\eta$ will be elevated towards the detection limit where SEDs will become less constrained (especially if there is sparser multi-wavelength coverage; e.g., non-detections and poor S/N) increasing uncertainties in the photo-$z$ fitting. Farmer is found to best perform at fainter magnitudes compared to Classic as first reported in \citet{Weaver2022}. Comparing \textit{Chandra}/COSMOS and COSMOS2020 Classic (X-ray), we find that the former has smaller $\sigma_\textrm{NMAD}$, which would imply less scatter. However, we find the outlier fractions are higher at $i < 22$ mag for \textit{Chandra}/COSMOS reaching $\sim 20\%$ at the faintest magnitude. From these tests, we can determine how reliable photo-$z$ measurements are for different subpopulations of galaxies.

Future surveys targeting COSMOS can easily use the compilation and run similar photo-$z$ validation tests as demonstrated above. Especially with the inclusion of \textit{JWST} photometry from large programs such as COSMOS-Web (\citealt{Casey2023,Franco2025,Harish2025,Shuntov2025}) and PRIMER (PI: J.~Dunlop), we may see improved photo-$z$ measurements especially given the wealth of near/mid-IR constraints that would break degeneracies such as the Lyman/Balmer Breaks that was noted in Figure \ref{fig:validation}. Overall, this compilation, and future versions as more spectroscopic redshifts are incorporated, will serve as a valuable legacy resource in order to calibrate and validate photo-$z$ measurements for future surveys.

\begin{figure}
	\centering
	\includegraphics[width=\columnwidth]{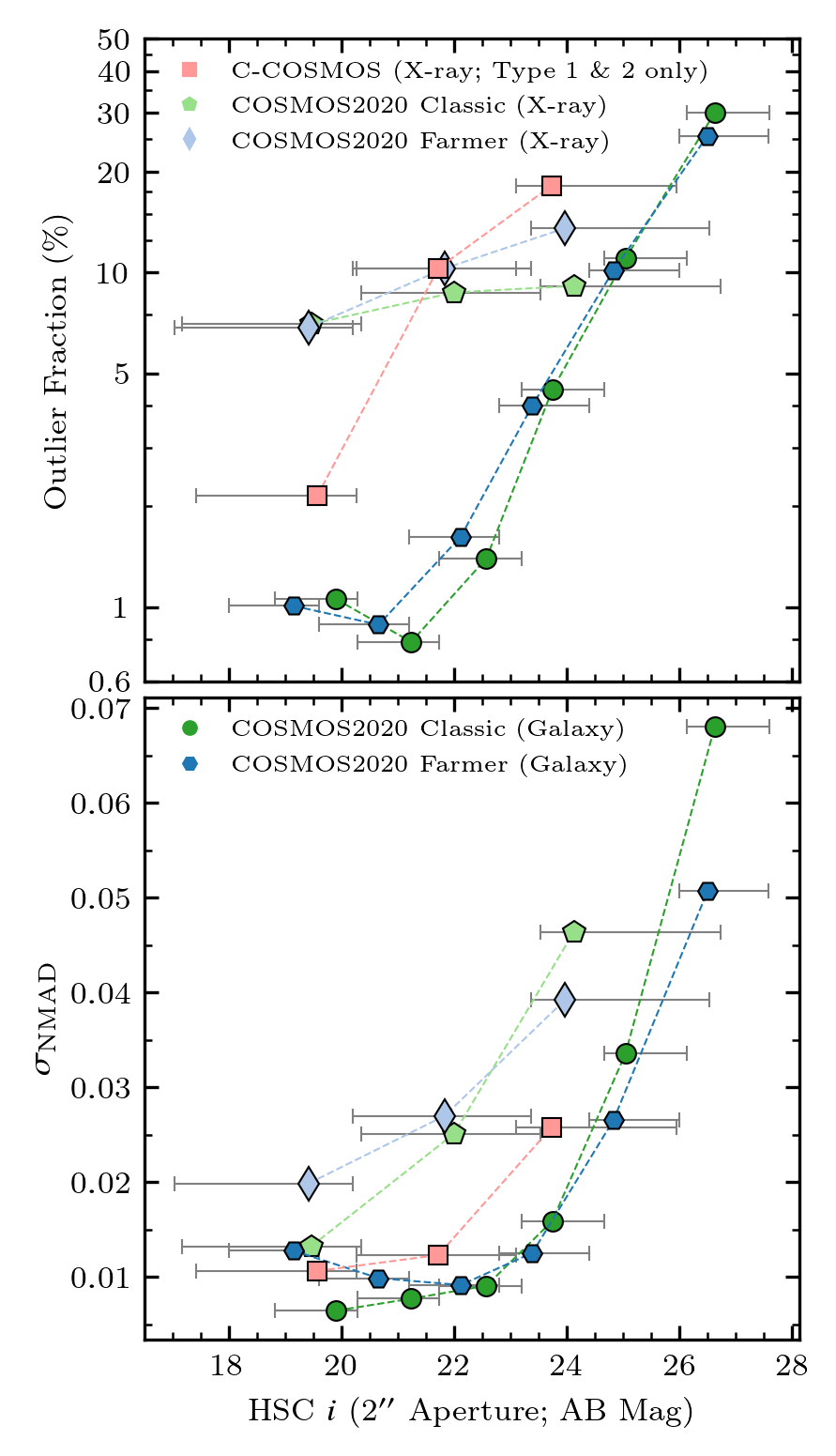}
	\caption{Photometric redshift quality assessment for different ranges of $i$ magnitude within $2''$ aperture for COSMOS2020 Classic, Farmer, and C-COSMOS (Type 1 \& 2 AGN only) using all reliable spec-$z$ measurements in the compilation. Outlier fractions (\textit{top panel}) and NMAD (\textit{lower panel}) are shown for each of the data sets. Both properties increase with fainter $i$ magnitude where $\sim 12\%$ of photo-$z$ measurements with associated spec-$z$ are invalidated by $\sim 25$ mag. This is also the magnitude range corresponding to high-$z$ candidates where contamination rates are higher, explaining the increase in outlier fraction and NMAD (e.g., LBG/\lya~emitter is found to be a low-$z$ interloper). As demonstrated here, the compilation can easily be used for high-quality photo-$z$ validation measurements, given the numerous spec-$z$ measurements, and can be subdivided into magnitudes or other properties allowing for a detailed assessment of photo-$z$ reliability in new samples.}
	\label{fig:photoz_validation_imag}
\end{figure}

\subsection{Self-Organizing Maps}
\label{sec:SOMs}

\begin{figure*}
	\includegraphics[width=\textwidth]{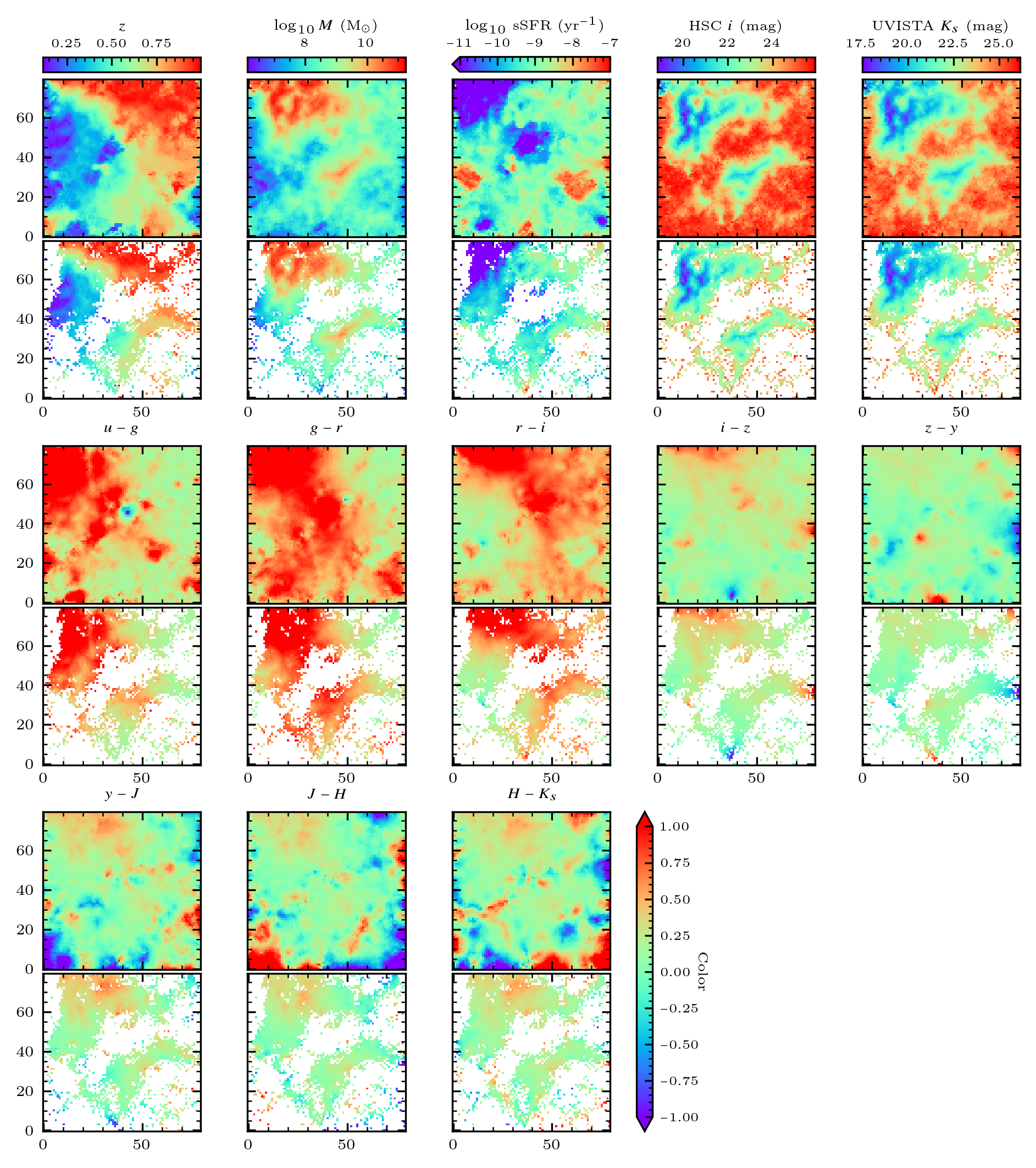}
	\caption{Low-$z$ ($0.1 < z < 1$) SOM with all the features used in the training shown in each panel. Only HSC $i$ and UltraVISTA $K_s$ magnitudes were not included in the training process and are shown here only for the purpose of identifying the magnitude ranges for each point in the SOM. Each \textit{odd row} shows the COSMOS2020 trained map. The spec-$z$ compilation is projected onto the SOM and is shown on each \textit{even row} and highlights the regions for which spectral coverage currently exists in COSMOS (based on the compilation) as well as regions for which there is no spectral coverage as of yet. Maps such as these are crucial in identifying sub-populations that have so far gone missing in past spectroscopic observations, such as the intermediate-mass, quiescent population ($x \sim 40$, $y \sim 45$), and the various high sSFR clouds that correspond to galaxies with strong line emission. Such visualizations of galaxy populations using all known photometric and spectroscopic datasets can provide valuable insight on how to proceed in the future in terms of observation strategy and target definitions.}
	\label{fig:SOM_lowz}
\end{figure*}

\begin{figure*}
	\includegraphics[width=\textwidth]{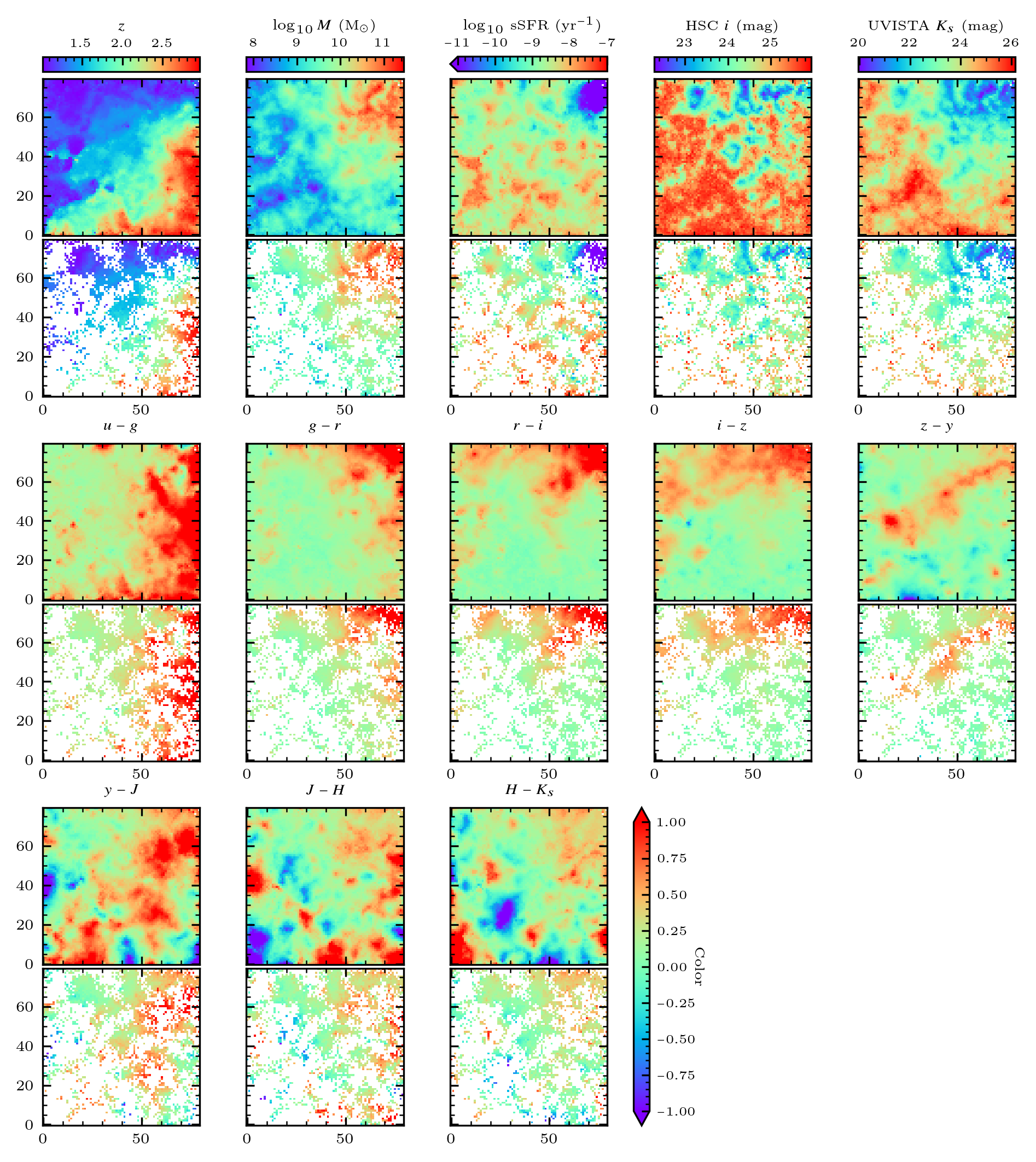}
	\caption{Mid-$z$ ($1 < z < 3$) SOM with similar description to Figure \ref{fig:SOM_lowz}.}
	\label{fig:SOM_midz}
\end{figure*}

\begin{figure*}
	\includegraphics[width=\textwidth]{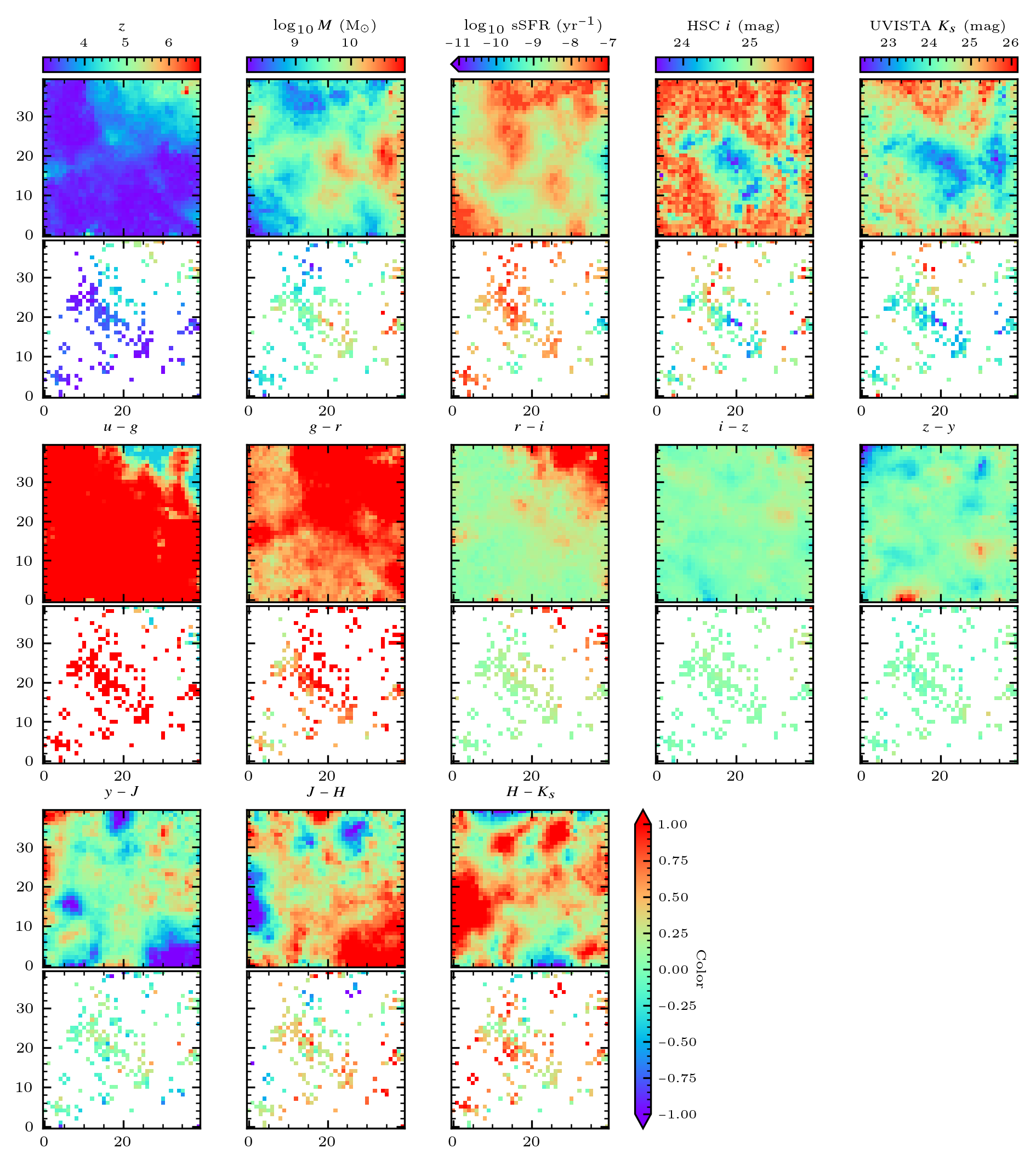}
	\caption{High-$z$ ($z > 3$) SOM with similar description to Figure \ref{fig:SOM_lowz}. One thing to note is the $u - g$ and $g - r$ colors which are dominated by our $2 \sigma$ upper limit cuts. This is why $u - g$ colors are constantly $> 1$ mag for the $z < 4$ population due to non-detection in $u$ and also why the $u - g$ colors are consistently blue at $z > 4$. This is not because the populations are blue rather that the $2 \sigma$ limits for CFHT/$u$ and HSC/$g$ are $28.1$ and $28.5$, respectively ($u -g \sim -0.4$ mag).}
	\label{fig:SOM_highz}
\end{figure*}

Self-organizing map (SOM; \citealt{Kohonen1990}) is an unsupervised machine learning approach that organizes higher dimensional datasets into a 2D organized grid space while also preserving topological relationships (e.g., data points mapped out into nearby points in the grid have similar features). This makes SOM quite powerful for identifying patterns within higher dimensional datasets. Past implementations of SOM have been used to primarily demonstrate implementation and optimization of template-based model fitting for photometric redshift and galaxy property determinations (e.g. \citealt{Geach2012,Masters2015,Hemmati2019a,Hemmati2019b,Sanjaripour2024,Abedini2025}). Such efforts are crucial in preparing the infrastructure for next-generation surveys planned with \textit{Euclid}, \textit{Roman}, and ELTs that will observe significantly large number of galaxies making traditional photometric redshift and physical property determination methods computationally inefficient.

For the scope of this work, we use SOMs to rather understand what subpopulations of galaxies currently have spectroscopic coverage within the compilation and which galaxy types are currently lacking any past observation (e.g., \citealt{Masters2017}). In this way, we can use archival datasets to get a big picture representation of what galaxy subpopulations currently are missing spectroscopic follow-up, thereby informing future survey strategies. This builds on a similar approach \citep{Masters2015} that led to programs such as C3R2 \citep{Masters2017} that used SOMs as the basis of their targeting strategy. Future programs such as 4MOST-4C3R2 \citep{Gruen2023} will also use a SOM-based strategy to target under-represented populations.

\subsubsection{Training the Maps}

We start by using the python-based \texttt{sompy} package \citep{moosavi2014sompy} to train the self-organizing maps using the COSMOS2020 Classic catalog limited to $i < 26$ mag, which takes into account the incompleteness limits of both our spectroscopic compilation and the Classic catalog. The training datasets are subdivided into three redshift bins such that we have three different SOMs: low-$z$ ($0.1 < z < 1$), mid-$z$ ($1 < z < 3$), and high-$z$ ($3 < z < 9$). Given that the low-$z$ population is so numerous, its features could potentially wash out those associated with the mid- and high-$z$ populations during the training process.

The features used in each SOM are the color combinations: $u - g$,~$g - r$,~$r - i$,~$i - z$,~$z - y$,~$y - J$,~$J - H$,~and $H - K_s$ and the photometric redshift (\texttt{lp\_zbest}), stellar mass (\texttt{lp\_mass\_best}), and sSFR (\texttt{lp\_sSFR\_best}). The last three are typically not included in SOM implementations; however, our main focus is not to use SOM to measure these three properties but rather use this information to understand which populations in terms of their SED shapes (colors), stellar mass, redshift, and sSFR are missing from past COSMOS spectroscopic observations. Each feature is normalized prior to the training process by its variance.
 
One main problem with SOM is the handling of missing data (e.g., non-detections). One approach is applying a detection threshold criterion (e.g., $> 5\sigma$ depth) to ensure that there is no missing data; however, this would severely limit the sample size as well as place an upper limit in redshift ($z < 3$; the Lyman Limit starts to fall into $u$ resulting in non-detections). Other approaches include linear interpolation of the data (e.g., detection in $z$ and $J$ but missing $Y$). \cite{Chartab2023} and \cite{LaTorre2024} investigated how manifold learning can be used to handle missing data within photometric catalogs; however, these approaches are primarily focused on improving physical property measurements using SOMs. 

Since our main objective is to assess which populations do/do not have spectroscopic coverage, this simplifies our case compared to past approaches in handling missing data. In our specific case, $z > 3$ populations lack detection as the Lyman limit shifts to redder passbands such that including estimated data via imputation will not provide any key physical information as it is expected these populations have no detections bluewards of the Lyman Limit due to the IGM transmission \citep{Madau1995,Inoue2014}. At $z < 3$, missing data is primarily associated with situations where the source was fainter than the detection threshold associated with a specific observed passband. As such, we assign the $2\sigma$ upper detection limit as measured in \cite{Weaver2022} where we have a non-detection or observed photometry fainter than the $2\sigma$ limit. The latter is to take into account data that may have increased photometric uncertainties.

The SOM maps are trained within a grid of $80 \times 80$ for the low-$z$ and mid-$z$ maps and a $40 \times 40$ grid for the high-$z$ map (accounting for the lower number of sources). We tested other grid sizes following the approach of \cite{Davidzon2019}. We start with a $30 \times 30$ grid space and train the SOM monitoring both quantization error, computational time, and average occupation. We then increment by 10 on each dimension until $100 \times 100$. We determined that our adopted grid sizes resulted in a minimized quantification error of 0.97, 0.99, and 1.36 for the low-$z$, mid-$z$, and high-$z$ maps, respectively, while also not heavily impacting the computational time. The maps are trained assuming 1000 iterations in the rough training and 1000 iterations in the fine-tuning training.

Figures \ref{fig:SOM_lowz} -- \ref{fig:SOM_highz} show our trained COSMOS2020 Classic SOM (\textit{odd row panels}) and our spec-$z$ compilation limited to \qf~$\sim 3 - 4$ projected within the trained maps (\textit{even row panels}) for all features. The trained maps inform us about which galaxy properties are currently photometrically observed within COSMOS while the projection of the spec-$z$ compilation provides insights on the subpopulations of galaxies that are currently observed and which have no spectroscopic coverage, at least for the COSMOS field and programs which make up the compilation. Future versions of the compilation that include new datasets and \textit{JWST}/NIRSpec and NIRCam grism redshifts will help fill in these gaps.

\subsubsection{Low-$z$ ($0.1 < z < 1$) Map}

One obvious population that we are missing across all redshifts is low-mass, faint systems. This is not surprising, as observing such sources, especially with ground-based observatories, is difficult and we have already shown how the spectroscopic completeness of the compilation drops significantly approaching $i \sim 26$ mag (see \S\ref{sec:completeness}). However, other distinct populations are also missing from the compilation, for which we provide the $(x, y)$ coordinates of the 2D maps as reference. Figure \ref{fig:SOM_lowz} shows the low-$z$ maps where we find a population of $0.1 < z< 1$ galaxies with $10^{8 - 9}$ \msol~and sSFR$ < 10^{-10}$ yr$^{-1}$ indicative of dwarf-like, quiescent populations ($x \sim 40, y \sim 45$). Although they are quite faint in $i$ ($\sim 25$ -- $26$ mag), their $K_s$ magnitudes are quite bright within the range of $22$ to $25$ mag. The color panels also show red colors for $u - g$, $g - r$, and $r - i$ indicating a rising SED followed by $\sim 0$ mag colors at redder passbands indicating a flat SED. The $u - g$ colors for a smaller subset of this region ($x \sim 42$, $y \sim 45$) suggest a bump in its SED at bluer wavelengths. On closer inspection, this is found to be due to photometric uncertainties near the $2\sigma$ detection limit in both the CFHT/$u$ and HSC/$g$ photometry.

Two other populations of low-mass, quiescent galaxies are noted at ($x \sim 15$, $y \sim 5$) and ($x \sim 75$, $y \sim 5 - 10$). The fact that these are organized in different regions by SOMs means that certain features were identified that made these populations separate from the main low-mass quiescent population at ($x \sim 40, y \sim 45$). Inspecting the color suggests a drop in $J$ flux density for ($x \sim 15$, $y \sim 5$)  and $H$ flux density for ($x \sim 75$, $y \sim 5 - 10$) followed by bright redder passbands. These populations of low-mass, low-$z$ quiescent galaxies could indicate populations that may be undergoing environmental quenching at $z < 1$ (e.g., \citealt{Peng2010,Darvish2016,Kawinwanichakij2017,Chartab2020}). Given their red colors, it is also important to investigate whether these sources are dusty systems mimicking the SEDs of quiescent galaxies. Identifying these populations via SOMs would allow the design of a potential survey program to target these regions based on their common features and ascertain the true nature of these sources.

Another population of quiescent galaxies missed within the compilation is found in the top-left corner of the map ($x <5$, $y > 60$), which corresponds to $z < 0.5$ galaxies exhibiting sSFR $< 10^{-11}$ yr$^{-1}$ with stellar masses of $>10^8$ \msol~and extending up to $10^{11}$ \msol. This population has $u - g$ and $g - r$ colors similar to other quiescent populations that are spectroscopically observed; however, its $r - i$ colors are significantly bluer, suggesting a flattening in its SED occurring at bluer bandpasses.

Figure \ref{fig:SOM_lowz} also shows 4 distinct populations of high sSFR ($> 10^{-8}$ yr$^{-1}$) not spectroscopically observed. In fact, the \textit{second row} clearly highlights how past spectroscopic follow-up programs in COSMOS have not significantly targeted sSFR $>10^{-8}$ yr$^{-1}$ systems. The 4 main distinct regions are located at ($x \sim 10$, $y \sim 30$), ($x \sim 35$, $y \sim 5$), ($x \sim 65$, $y \sim 25$), and ($x \sim 75$, $y \sim 35$). The clumps have different redshift ranges with ($x \sim 10$, $y \sim 30$) and ($x \sim 35$, $y \sim 5$), consisting of $ < 10^8$ \msol~and $z \sim 0.1 - 0.5$ populations. The other two clumps consist of $z > 0.7$ galaxies with stellar masses of $\sim 10^{8 - 9}$ \msol. The colors of each region are unique as well where, for example, the ($x \sim 75$, $y \sim 35$) clump is strongly red in $i - z$ and blue in $z - y$ colors indicative of high EW emission line contribution in HSC $z$ which at $z \sim 0.8$ would correspond to \oiii4959,5007\AA. Clump ($x \sim 35$, $y \sim 5$) also shows strongly red $r - i$ and $z - y$ and blue $i - z$ colors which at $z \sim 0.25$ corresponds to high \halpha~EW emission line contribution in HSC $i$. 

These high sSFR clumps trace extreme emission line galaxy (EELG) populations that have been studied in the past as local analogs of reionization-era galaxy populations (e.g., \citealt{Cardamone2009,Amorin2015,Yang2017,Izotov2021,Khostovan2025}). However, such populations are also quite low in number density as indicated in past studies of emission line EW distributions (e.g., \citealt{Fumagalli2012,Sobral2014,Faisst2016,Khostovan2016,Khostovan2021,Khostovan2024}), which may also explain the lack of spectroscopic coverage of these sources within COSMOS. Using SOMs coupled with the compilation, we can inform future survey strategies on how to identify distinct EELG populations for follow-up spectroscopic observations.

\subsubsection{Mid-$z$ ($1 < z < 3$) Map}

As shown in Figure \ref{fig:SOM_midz}, past spectroscopic coverage is spread out throughout the map; however, there are still key gaps in spectroscopic follow-up programs of different mid-$z$ galaxy populations. For example, we find a clear concentration of $1 < z < 2$ quiescent galaxies (low sSFR) with stellar mass $\sim 10^{9.5 - 11.4}$ \msol~and rising SEDs as indicated by the color panels in the \textit{top right} portion of the trained SOM. The compilation does cover a portion of this population limited to $z \sim 1$ galaxies at primarily $> 10^{10.5}$ \msol~but we are missing the higher redshift quiescent galaxies and lower mass populations that make up this full region. Such quiescent sources are ideal candidates to investigate massive galaxy growth and quenching mechanisms (\citealt{Chartab2025}). We do note that spectroscopic follow-up could also reveal a fraction of this population are dusty, star-forming galaxies at low-$z$ that mimic the SED shape of high-$z$ quiescent galaxies. 

Other subpopulations include the regions at ($x < 40$, $y < 40$) that are not as well populated compared to other parts of the map. Populations residing here are $<10^{9.5}$ \msol~galaxies with sSFR ranging from normal ($\sim 10^{-9}$ yr$^{-1}$) to bursty ($\gtrsim 10^{-8}$ yr$^{-1}$) star-forming galaxies. As indicated in the \textit{even row} panels highlighting the spec-$z$ compilation project on the trained SOM, there are several points that probe these regions but only tend to cover a single grid point such that the full sub-population around the associated regions are not well probed. These key regions could give us insight into normal \& bursty star-forming conditions within low-mass galaxies at the cosmic peak of star-formation activity.

\subsubsection{High-$z$ ($z > 3$) Map}

Figure \ref{fig:SOM_highz} shows the high-$z$ SOM where the spec-$z$ projection on the map unsurprisingly highlights the lack of spectroscopic observations covering this period of cosmic time. This is especially evident in the \textit{top right} corner, which is populated by $z > 4$ galaxies with the projection maps showing a clear lack of spectroscopic coverage. We note this will quickly change as \textit{JWST} NIRSpec and NIRCam WFSS programs publish redshifts, which will be later included in the compilation. The main reason for much of the empty space is due to the observationally faint nature of high-$z$ galaxy populations.

The COSMOS2020 SOM highlights key regions of massive ($>10^{10.5}$ \msol) galaxy populations at ($x \sim 30 - 40$, $y \sim 11 - 22$) and ($x \sim 20 - 25$, $y \sim 19 - 21$) that correspond to $3 < z < 5$ galaxies. These systems are uniquely massive for their redshifts with sSFR $\sim 10^{-9}$ yr$^{-1}$ and colors indicating a rising SED towards redder wavelengths suggesting that these may be populations of dusty, massive star-forming galaxies. Spectroscopic confirmation is needed to validate the nature of these sources; however, this can provide a great opportunity to study both dust properties and rapid galaxy growth at relatively high-$z$.

\section{Future of the Compilation}\label{sec:discussion}
\label{sec:future}
This spectroscopic redshift compilation will be a continuously evolving data product maintained by the COSMOS Collaboration with legacy value for the astronomical community. The infrastructure is already set within our compilation pipeline as described in \S\ref{sec:compilation_pipeline} for the inclusion of new public spectroscopic datasets as they are made available. The future includes exciting programs that will provide a wealth of spectroscopic redshifts. 

Other programs expected to be part of the compilation in the near future include various \textit{JWST} spectroscopic programs as data starts to become public and redshifts are published. These datasets will heavily populate the high-$z$ portion of the compilation and also extend the compilation to fainter populations. \textit{JWST}/NIRCam WFSS slitless spectroscopy will also provide a wealth of redshifts to the compilation. One major slitless spectroscopy program is COSMOS-3D, which will cover $\sim 0.33$ deg$^2$ of COSMOS-Web and observe $\sim 20000$ galaxies and $\sim 5000$ AGN with $\sim 4000$ galaxies and $\sim 500$ AGN expected to be observed at $z > 5$ populating the high-$z$ end of the compilation. 

Future next-gen space-based missions and ground-based observatories will provide a wealth of redshifts that will be included in the compilation when made available. COSMOS is one of six \textit{Euclid} Auxiliary Fields that will provide slitless grism spectroscopic coverage of the full 2 deg$^2$ field down to $2 \times 10^{-16}$ erg s$^{-1}$ cm$^{-2}$ ($3.5\sigma$; \citealt{Scaramella2022}). \textit{Roman} is expected to launch in May 2027 and will observe numerous emission line galaxies with grism spectroscopy in a single $0.28$ deg$^2$ pointing (e.g. \citealt{Valentino2017,Zhai2019,Wang2022,Khostovan2024}). Subaru Prime Focus Spectrograph (PFS) Galaxy Evolution Survey will measure $\sim 2400$ redshifts per $1.25$ deg$^2$ field-of-view, with coverage from 0.38 -- 1.26 \micron~with COSMOS being one of the targeted fields \citep{Greene2022}. The VLT/MOONS Redshift-Intensive Survey Experiment (MOONRISE) will target 1 deg$^2$ of COSMOS as one of its three main fields with $\sim 1000$ redshifts per $0.14$ deg$^2$ field-of-view \citep{Maiolino2020}. WAVES/4MOST will observe COSMOS as one of its Deep Drilling Fields covering $\sim 4$ deg$^2$ in a single pointing observing 45000 sources within 0.37 -- 0.95 \micron~\citep{Driver2019}.

It is expected, as described above, that the compilation will grow significantly given the volume of data expected to be available in the near future. As such, the compilation will have periodic data releases made on a yearly basis and will be available via our \texttt{GitHub} repository: \href{https://github.com/cosmosastro/speczcompilation}{ cosmosastro/speczcompilation}, facilitating quick access to the latest catalogs.

\section{Summary \& Conclusions}
\label{sec:conclusions}

We present the first data release of the COSMOS Spectroscopic Redshift Compilation that represents $\sim20$ years worth of spectroscopic studies within this single legacy field. The work encompasses not just observations and subsequent analysis by various groups, but also reflects the many years of work gathering and compiling all published datasets into a single, concise, and uniform data product. Our main results are as follows:

\begin{enumerate}[label=(\roman*)]
    \item Redshifts from \nprograms~programs were gathered via mining catalog database systems (e.g., IRSA, ESO Science Portal, CDS/VizieR), survey websites, publications, and direct contact with PIs. The compilation covers 10 deg$^2$ centered on COSMOS (2 deg$^2$ legacy field) in preparation for next-generation wide-area surveys. A total of \allSources~redshifts for \uniqueSources~unique sources make up the total compilation across all assigned quality flags. 
    \item We designed a compilation pipeline that gathers all our identified data sets, corrects for varied astrometric calibrations on a program-by-program basis, converts quality assessment flags to our uniform system for each individual data set, and compiles it all in a master catalog. This catalog is then used to identify and flag duplicates (e.g., sources that have multiple redshift measurements). The pipeline is also set up in such a way that new data can be efficiently appended to the compilation allowing for periodic updates.

    \item Included in the compilation are physical properties measured from \cigale~and \lephare~SED fitting using the best spectroscopic redshift per source and limited to the high \qf~subset. 

    \item The compilation is complete at the 50\% level down to $i \sim 23.4$ and $K_s \sim 21.6$ mag and is most spectroscopically complete towards the central regions of COSMOS corresponding to the CANDELS region, which is expected given the number of spectroscopic programs covering this one key area.

    \item We find the compilation has a median stellar mass that varies with redshift. At $z < 0.4$, the median stellar mass rises from $10^{8.4}$ \msol~to $\sim10^{10}$ \msol~by $z \sim 0.6$ and then decreases up to $z \sim 7$ reaching a median stellar mass of $\sim10^{9.4}$ \msol. 
    
    \item The stellar mass distributions suggest a bimodal population of low-mass and intermediate mass star-forming galaxies and massive, quiescent galaxies up to $z \sim 1.25$. Rest-frame $NUVrJ$ colors also highlight both the quiescent and star-forming populations as well as how the population progressively shifts towards the blue loci associated with dust-poor, young population-dominated star-forming galaxies at $z > 2$. 

    \item The SFR -- stellar mass correlation shows how our compilation aligns well with past measurements as well as the turnoff that occurs at $>10^{10}$ \msol~associated with quenching populations. At $z > 3$, we find that the compilation shifts towards galaxies with high sSFR ($> 10$ Gyr$^{-1}$) and 10 - to - 100 Myr SFR ratios $> 1$ indicating recent and potentially bursty star-formation activity.

    \item The compilation covers a diverse range of environments where we demonstrate how the compilation can be used to visualize the multi-component structure of large overdense regions such as `Hyperion' (3 overdensities) and a $z \sim 1.25 - 1.26$ structure (2 overdensities).

    \item We demonstrate how the compilation can be used for validating photometric redshifts in galaxy survey programs. We used several versions of the COSMOS catalogs and demonstrated how we can get further information by looking at photometric redshift calibration metrics (e.g., outlier fractions, $\sigma_\textrm{NMAD}$) as dependent on other properties such as HSC $i$ magnitudes. 


    \item We demonstrate how self-organizing maps can be used to gain valuable insight into how well we are spectroscopically covering various galaxy subpopulations. Surprisingly, our coverage of the $z < 1$ population is still incomplete where we are missing a population of intermediate-mass quiescent galaxies as well as high sSFR systems. Other populations include $z \sim 2$ massive quiescent galaxies and $< 10^{9.5}$ \msol~high sSFR systems. Overall, SOMs combined with our spec-$z$ compilation can provide us with valuable insight into identifying which populations we require to get a complete spectroscopic view of galaxy populations thereby informing future observing strategies. 
\end{enumerate}

The future of astronomy will be heavily dominated by the introduction of new and large data volumes coming from facilities such as \textit{Roman}, \textit{Euclid}, Subaru/PFS, VLT/MOONS, and many others. It is therefore of great importance to keep track of not only what new data is coming in, but also what data we currently have on hand. Data products such as the COSMOS Spectroscopic Redshift Compilation are crucial legacy resources for the community that will be periodically updated to keep track of the latest redshift measurements. 

\section*{acknowledgments} 
The authors thank the anonymous referee for their comments and suggestions that aided in enhancing the quality of this work. The authors acknowledge the hard work of all members of each respective program that have made this compilation possible. This material is based upon work supported by the National Science Foundation under Grant No. 2009572 and by NASA under award No. 80NSSC22K0854. The French contingent of the COSMOS team is partly supported by the Centre National d’Etudes Spatiales (CNES). OI acknowledges the funding of the French Agence Nationale de la Recherche for the project iMAGE (grant ANR-22-CE31-0007). This research has made use of the VizieR catalogue access tool, CDS, Strasbourg, France \citep{Ochsenbein2000}. This research has made use of NASA’s Astrophysics Data System.

Some of the data presented herein were obtained at Keck Observatory, which is a private 501(c)3 non-profit organization operated as a scientific partnership among the California Institute of Technology, the University of California, and the National Aeronautics and Space Administration. The Observatory was made possible by the generous financial support of the W. M. Keck Foundation. The authors wish to recognize and acknowledge the very significant cultural role and reverence that the summit of Maunakea has always had within the Native Hawaiian community. We are most fortunate to have the opportunity to conduct observations from this mountain.

Based on observations obtained at the international Gemini Observatory, a program of NSF NOIRLab, which is managed by the Association of Universities for Research in Astronomy (AURA) under a cooperative agreement with the U.S. National Science Foundation on behalf of the Gemini Observatory partnership: the U.S. National Science Foundation (United States), National Research Council (Canada), Agencia Nacional de Investigaci\'{o}n y Desarrollo (Chile), Ministerio de Ciencia, Tecnolog\'{i}a e Innovaci\'{o}n (Argentina), Minist\'{e}rio da Ci\^{e}ncia, Tecnologia, Inova\c{c}\~{o}es e Comunica\c{c}\~{o}es (Brazil), and Korea Astronomy and Space Science Institute (Republic of Korea).


\facilities{AAT, \textit{HST} (ACS, WFC3-IR), ALMA, LMT, SMA, NOEMA, WHT (LIRIS), TNG (NICS), Keck:I (LRIS, MOSFIRE), Keck:II (DEIMOS, NIRSPEC), LBT (LUCI1, LUCI2), VLT:Antu (FORS1, FORS2, KMOS), VLT:Kueyen (X-SHOOTER) VLT:Melipal (VIMOS), VLT:Yepun (MUSE, SINFONI), MMT (Binospec, Hectospec), Mayall (DESI), Subaru (FMOS, FOCAS, MOIRCS), Gemini:South (GMOS), HET (VIRUS), Magellan:Baade (IMACS, FIRE), Magellan:Clay (LDSS3, M2FS), \textit{Spitzer}:IRS, \textit{Gaia}, Sloan, Du Pont}

\software{\texttt{astropy} \citep{astropy:2013, astropy:2018, astropy:2022}, \cigale~\citep{Boquien2019}, \lephare~\citep{Arnouts1999,Ilbert2006}, 
          \texttt{numpy} \citep{harris2020array}, \texttt{sompy} \citep{moosavi2014sompy}
}

\bibliography{main.bib}{}
\bibliographystyle{aasjournal}

\appendix

\section{Column Definitions of Compilation Files}

\begin{table*}
\caption{Description of all columns within the spec-$z$ compilation FITS files. }
\label{table:comp}
\begin{tabular*}{\textwidth}{@{\extracolsep{\fill}}llcl}
    \hline
    Column & Format & Units & Description \\
    \hline
    \texttt{Id\_specz}		    &  long    &      & Assigned Unique Identification in the Compilation \\
    \texttt{Id\_original}	   	&  string  &      & Original Identification of the source \\
    \texttt{ra\_original}	   	&  float   & deg  & Original Right Ascension\\
    \texttt{dec\_original}	  	&  float   & deg  & Original Declination \\
    \texttt{ra\_corrected}	  	&  float   & deg  & Astrometry Corrected	Right Ascension \\
    \texttt{dec\_corrected}	 	&  float   & deg  & Astrometry Corrected	Declination \\
    \texttt{Priority}		    &  float   &      & Flag 1: Use this redshift, especially in case of duplicate sources \\
    \texttt{specz}  		  	&  float   &      & Spectroscopic Redshift \\
    \texttt{flag} 			    &  long    &      & Quality Assessment Flag (see \S\ref{sec:QF_scheme}) \\
    \texttt{Confidence\_level}	&  long    &      & Confidence Level (see \S\ref{sec:QF_scheme}) \\
    \texttt{survey}			    &  integer &      & Survey ID in \texttt{\_surveys.list} and Table \ref{table:datasets}  \\
    \texttt{compilation\_year}	&  long    &      & Past versions of the compilation where these sources were introduced	 \\
    \texttt{Id\_COS20\_Classic} &  long    &      & COSMOS2020 Classic Identification \\
    \texttt{ra\_COS20\_Classic}	&  float   & deg  & COSMOS2020 Classic Right Ascension  \\
    \texttt{dec\_COS20\_Classic}&  float   & deg  & COSMOS2020 Classic Declination  \\
    \texttt{Id\_COS20\_Farmer}	&  long    &      & COSMOS2020 Farmer Identification \\
    \texttt{ra\_COS20\_Farmer}	&  float   & deg  & COSMOS2020 Farmer Right Ascension \\
    \texttt{dec\_COS20\_Farmer}	&  float   & deg  & COSMOS2020 Farmer Declination 	 \\
    \texttt{Id\_COSMOS25}       &  long    &      & COSMOS2025 Identification \\
    \texttt{ra\_COSMOS25}       &  float   & deg  & COSMOS2025 Right Ascension \\
    \texttt{dec\_COSMOS25}      &  float   & deg  & COSMOS2025 Declination \\
    \texttt{Id\_COSMOS15} 	    &  long    &      & COSMOS2015 Identification \\
    \texttt{ra\_COSMOS15}	    &  float   & deg  & COSMOS2015 Right Ascension  \\
    \texttt{dec\_COSMOS15}	    &  float   & deg  & COSMOS2015 Declination 	 \\
    \texttt{Id\_COSMOS09} 	    &  long    &      & COSMOS2009 $i$-band Identification \\
    \texttt{ra\_COSMOS09}	    &  float   & deg  & COSMOS2009 $i$-band Right Ascension  \\
    \texttt{dec\_COSMOS09}	    &  float   & deg  & COSMOS2009 $i$-band Declination 	 \\
    \texttt{photoz} 		    &  float   &      & COSMOS2020 Classic Photo-z LePhare  \\
    \texttt{photoz\_type}	    &  long    &      & 0: Galaxy; 1: star; 2: Xray source; -9: failure in fit \\
    \texttt{GroupID}            &  long    &      & ID corresponding to groups of redshifts per unique source (no duplicates $= -1$)\\
    \texttt{GroupSize}          &  long    &      & Number of duplicate redshifts in a given group\\
    
    \hline
\end{tabular*}
\end{table*}
\end{document}